\documentclass[sigconf,dvipsnames]{acmart}

\usepackage{latexsym}
\usepackage{graphicx}
\graphicspath{{./images/}}
\usepackage{booktabs} % for formal tables
\usepackage{amsmath}  % for aligning equations
\usepackage{subcaption}
\usepackage{caption}
\usepackage{tikz}
\usepackage{colortbl} % for color in tables
\usepackage{framed}
\usepackage{multirow}
\usepackage{multicol}
\usepackage{hyperref}
\usepackage{url}
\usepackage{balance}
\usepackage{verbatim}
\usepackage{cancel}
\usepackage{xspace} % for correcting space after macro commands
\usepackage[ruled,vlined]{algorithm2e}
\usepackage{bbold} % for writing mathbb{1}
\usepackage{balance}

\newcommand{\struct}[1]{\texttt{\small #1}}
\newcommand{\utterance}[1]{\textit{#1}}
\newcommand{\phrase}[1]{\textit{``#1''}}

\newcommand {\anecdote}[1]{\colorbox[RGB]{236,236,236}{\parbox{8cm}{{#1}}}}
\newcommand {\itemslist}[1]{$\langle$\struct{#1}$\rangle$}

\newenvironment{Snugshade}[1][236,236,236]{
    \setlength{\itemsep}{0pt}
    \setlength{\parsep}{0pt}
    \setlength{\topsep}{0pt}
    \setlength{\partopsep}{0pt}
    \setlength{\leftmargin}{1.5em}
    \setlength{\labelwidth}{0em}
    \setlength{\labelsep}{0em} 
    \setlength{\parskip}{0pt}
    \definecolor{shadecolor}{RGB}{#1}%
    \begin{snugshade}
}{%
    \end{snugshade}%
}

\newcommand{\clocq}{\textsc{Clocq}\xspace}

\setcopyright{acmcopyright}
\copyrightyear{2022}
\acmYear{2022}
\acmConference[WSDM'22]{The 15th ACM International Conference on Web Search and Data Mining}{February 2022}
{Phoenix, Arizona}
\acmYear{2022}
\copyrightyear{2022}
\begin{document}

\title{Beyond NED: Fast and Effective Search Space Reduction for Complex Question Answering over Knowledge Bases}

\author{Philipp Christmann}
\affiliation{%
  \institution{Max Planck Institute for Informatics\\Saarland Informatics Campus}
  \streetaddress{Saarland Informatics Campus}
  \city{Saarbruecken}
 \country{Germany}
}
\email{pchristm@mpi-inf.mpg.de}

\author{Rishiraj Saha Roy}
\affiliation{%
  \institution{Max Planck Institute for Informatics\\Saarland Informatics Campus}
  \streetaddress{Saarland Informatics Campus}
  \city{Saarbruecken}
    \country{Germany}
}
\email{rishiraj@mpi-inf.mpg.de}

\author{Gerhard Weikum}
%\authornote{Dr.~Trovato insisted his name be first.}
\affiliation{%
  \institution{Max Planck Institute for Informatics\\Saarland Informatics Campus}
  \streetaddress{Saarland Informatics Campus}
  \city{Saarbruecken}
    \country{Germany}
}
\email{weikum@mpi-inf.mpg.de}

\renewcommand{\shortauthors}{P. Christmann et al.}

\newcommand{\squishlist}{
    \begin{list}{$\bullet$}{ 
        \setlength{\itemsep}{0pt}
        \setlength{\parsep}{1pt}
        \setlength{\topsep}{1pt}
        \setlength{\partopsep}{0pt}
        \setlength{\leftmargin}{1.5em}
        \setlength{\labelwidth}{1em}
        \setlength{\labelsep}{0.5em} 
    } 
}

\newcommand{\squishend}{
  \end{list}  }
  
\newcommand{\floor}[1]{\lfloor #1 \rfloor}

\newcommand{\GW}[1]{{\color{blue}{GW: #1}} }
\newcommand{\PC}[1]{{\color{orange}{PC: #1}} }

\newcommand{\myparagraph}[1]{\noindent \textbf{#1}.}

\setcounter{secnumdepth}{4}

% !TEX root = ../2021-sigir-sp-clocq.tex
\begin{abstract}

% comment-free abstract for easy easychair
Answering complex questions over knowledge bases (KB-QA) faces huge input data with billions of facts, that involve millions of entities and thousands of predicates. For efficiency, QA systems first reduce the answer search space by identifying a set of facts that is likely to contain all answers and relevant cues. The most common technique for doing this is to apply named entity disambiguation (NED) systems to the question, and retrieve KB facts for the disambiguated entities. This work presents CLOCQ, an efficient method that prunes irrelevant parts of the search space using {KB-aware signals}. CLOCQ uses a top-$k$ query processor over score-ordered lists of KB items that combine signals about lexical matching, relevance to the question, coherence among candidate items, and connectivity in the KB graph. Experiments with two recent QA benchmarks for complex questions demonstrate the superiority of CLOCQ over state-of-the-art baselines with respect to answer presence, size of the search space, and runtimes.

\end{abstract}





\maketitle
% !TEX root = ../2021-sigir-sp-clocq.tex
\section{Introduction}
\label{sec:intro}

\myparagraph{Motivation} 
Large 
%curated 
knowledge bases (KBs) like Wikidata~\cite{vrandevcic2014wikidata}, DBpedia~\cite{auer2007dbpedia}, YAGO~\cite{suchanek2007yago}, and Freebase~\cite{bollacker2008freebase}
% DBpedia, Yago or Wikidata
are ideal sources for answering factual questions that have
crisp entity lists as answers
\cite{bast2015more,reinanda2020knowledge,yih2015semantic,abujabal2018never,huang2019knowledge,berant2013semantic,yahya2012natural,yih2015semantic}.
% loosely define facts upfront, not just triples
Such KBs are comprised of \textit{facts}, structured as \struct{<subject, predicate, object>} triples, often augmented with
% context information represented as
qualifier predicates and objects for context~\cite{pellissier2016freebase,galkin2020message,hernandez2015reifying,leeka2016quark}.
% need one line to motivate complex QA
Answering complex questions with multiple entities and predicates is one of the most actively researched topics in
QA over knowledge bases (KB-QA) today~\cite{saharoy2021question,qiu2020stepwise,he2021improving,vakulenko2019message,bhutani2019learning}, and this is the setting for this paper.
Systems answering such complex questions
%over KBs 
either build explicit structured queries~\cite{saharoy2021question,bhutani2019learning,luo2018knowledge,ding2019leveraging} or perform approximate graph search~\cite{lu2019answering,vakulenko2019message,sun2019pullnet} to arrive at the answer. %In either case, the input is i) a question, and ii) the entire KB with billions of facts. %
%The latter, in its entirety, can be rather unwieldy to process -- entity and predicate disambiguation becomes very challenging, and the processing time increases manifold. 
To this end, 
% most
systems learn models for mapping question words to
% canonicalized
KB items, where 
% Even if the search takes the form of
% an incremental exploration, driven by
% learning models, 
the huge size of the KB poses a stiff challenge.
Concretely, whole KBs are often more than 2 Terabytes
% on disk:
in size:
this makes the development of QA systems over them
% , especially for new entrants,
a rather daunting task.
%
%As a result, 
% Therefore,
Most KB-QA systems thus prune the search space for candidate answers using 
%off-the-shelf 
% methods % and tools
% for
Named Entity Disambiguation (NED). 

\myparagraph{Limitations of the state-of-the-art}
NED methods~\cite{van2020rel,ferragina2010tagme,hoffart2011robust,DBLP:journals/tkde/ShenWH15,li2020efficient,dubey2018earl} map mentions in questions 
(single words or short phrases)
to KB entities, and the QA system subsequently uses {only the facts containing these entities} as its \textit{search space} for locating the answer(s). 
%Key limitations of such general-purpose NED %~\cite{li2020efficient,blanco2015fast,ferragina2010tagme,van2020rel} 
%are:
However, general-purpose NED tools have major limitations in this context:
i) they are not tailored for downstream use by KB-QA systems;
ii) they usually disambiguate only named entities and disregard words and phrases that denote general concepts, types or predicates;
and iii) they typically output merely the top-1 entity per mention, missing out on further candidates that can serve as relevant cues.
Even methods designed for short input texts, like \textsc{Tagme}~\cite{ferragina2010tagme} and
\textsc{Elq}~\cite{li2020efficient}, have such limitations.

% touch upon key ideas in one sentence
\myparagraph{Approach} To address these
% limitations,
concerns,
we propose
% \clocq\footnote{\clocq stands for search space \underline{C}ontraction with \underline{L}ists and top-$k$ \underline{O}perators for \underline{C}omplex \underline{Q}uestion answering},
\clocq (\underline{C}ontract-ing answer spaces with scored \underline{L}ists and top-$k$ \underline{O}perators for \underline{C}omplex \underline{Q}A, pronounced \phrase{Clock}),
a time- and space-efficient
%  need to connect with clock analogy
method that operates over 
{\em all KB items} to produce top-$k$ candidates for entities, types, concepts and predicates.
%(for entities and predicates)
% ideally types and literals, but we don't have them in our example, and makes the writing a bit clumsy if we have to say E, P, T, L everywhere
%per question cue. 
\begin{comment}
\clocq stands for search space \underline{C}ontraction with
% index
\underline{L}ists and top-$k$ \underline{O}perators for \underline{C}omplex \underline{Q}uestion answering
% connect clock with sub-second completion times later
(with an intended wordplay on \textit{clock}, that symbolizes criticality of timing in QA algorithms), and runs in the following setup. 
\end{comment}
%GW: don't *explain* puns
%
Consider the 
%following 
complex question
% (
on the FIFA World Cup 2018:
%): 
%where it is clear to a human that the FIFA World Cup 2018 final is being referred to, but can be quite challenging for a machine:
\begin{Snugshade}
\begin{center}
    \utterance{Who scored in the 2018 final between France and Croatia?}
\end{center}
\end{Snugshade}
\noindent Most systems for complex KB-QA tackle the answering process in two phases~\cite{saharoy2021question}. 
First, they disambiguate question tokens to entities in the KB.
% via NED.
These entities
% question keywords (like \phrase{scored}, \phrase{2018}, \phrase{final}, \phrase{france}, \phrase{croatia}) to specific KB items (predicate: \struct{goal scored by} and entities: \struct{2018 FIFA WC final, France national football team, Croatia national football team}). This is done
% to
\textit{establish a reduced search space} for the QA system, that can either be
% \textit{hard} (
an \textit{explicit} set of facts containing these KB entities~\cite{vakulenko2019message,lu2019answering,sun2018open,sun2019pullnet},
or involve
% or \textit{soft} (
\textit{implicit} grounding to a small zone in the KB 
%created when these items are used to build 
% for
via
structured queries containing these entities~\cite{yih2015semantic,bhutani2019learning,lan2020query,yahya2012natural}.
Second, depending upon the approach in the first phase, KB-QA systems
either search for the answer in the retrieved facts,
or build a complex query in SPARQL-like syntax that would return the answer when executed.
% find the answer among a reasonably small set of candidates.
\clocq tries to improve
% Our approach is to tackle
the effectiveness and the efficiency of the first phase above.
% to create this search space for KB-QA systems. It follows that the core functionality of \clocq is to accurately disambiguate question tokens to KB-items, which is performed using top-$k$ algorithms.
Therefore, the output of \clocq is a small set of disambiguated KB items and facts containing these items, 
%that 
%directly serves as input to the QA model for phase ii).
% connect output of clocq to input of kbqa model
and this
% tractable subset of the KB
is fed into the
% answering
second
phase.
Answer presence in the KB subspace inherently sets an upper bound to the performance of the downstream KB-QA system,
making fast and effective search space reduction a vital step in the QA
% process.
pipeline.

\myparagraph{Method} \clocq first builds inverted lists of KB items per 
question
% term 
word
%(stopwords dropped)
% stoplists standard in IR
% no need to bring up bigrams, ner phrases here, keep it simple now
%using shallow lexical signals, with a focus on retaining as many candidates as possible.
with {\em term matching scores} based on
% BM25.
TF-IDF.
% \GW{BM25:correct? what about word2vec, BERT etc.?}
% Directly aggregating these lists to identify the top-$k$ from each is sub-optimal, as explained in the method and validated in the experiments. Rather,
Top-ranked items from these lists, up to a certain depth, are
% candidates that are 
then scored and ranked by a combination of 
%It then re-arranges items in these lists into \textit{item-wise score-ordered sequences}, where scores come from
\textit{global signals},
% \textit{set-level signals}
like semantic coherence
between items and
%among the top candidates and
connectivity in the KB graph,
% density
%in the corresponding knowledge graph,
and
% \textit{item-level signals}
\textit{local signals}
like 
%semantic 
relatedness to the question and 
%degree of term match. 
term-matching score.
%Fagin's 
These scoring signals are computed at question time: %,
% with quadratic-time cost, 
this is made feasible with \clocq's novel KB representation and storage model,
that substantially speeds up lookups with respect to existing solutions.
% but the implementation is very efficient and not a
% run-time
% runtime
% https://techterms.com/definition/runtime
% bottleneck.
% Fagin's
% A top-$k$
The {threshold algorithm} (TA)~\cite{fagin2003optimal} 
%~\cite{anh2006pruned,fagin2003optimal,theobald2008topx} 
is applied for 
extracting the top-$k$ candidates
% aggregating the scoring signals
for each question term separately.
%\textit{each set of lists}, and provides us with top-$k$ KB items \textit{per question term}. 
Since it may not always be obvious how to choose $k$ for every term, we also have an entropy-based mechanism for making this choice automatically.
The union of the per-term top-$k$ items forms 
%final disambiguation, 
a pool of relevant KB items,
and their KB facts
% (including qualifiers)
% are 
% above could be misleading, we don't really pass on in expts.
is the output of \clocq
that would be
passed on to the answering phase of a KB-QA system.
% as the output of \clocq.
% Concretely, systems can outsource this phase to \clocq, and directly proceed with the answer ranking (note that \clocq provides QA systems with both a reduced set of facts as well as the question-relevant KB items).
% something on the KB model, connect back to the clock analogy
%
%Note that
%search space reduction is a preprocessing task for a QA pipeline, and should be done as fast as possible: 
% We compute this reduction of the answer search space in sub-second times, using a novel KB index that takes a fact-centric view of the KB, efficient code for list construction and the threshold algorithm (TA)~\cite{fagin2003optimal}.
Experiments with two recent KB-QA benchmarks and a suite of NED-based competitors~\cite{ferragina2010tagme,hoffart2011robust,van2020rel,li2020efficient,dubey2018earl} show the benefits of \clocq: it obtains the highest \textit{answer presence} in the retained subset of the KB, with
tractable search space size and
% competitive
sub-second
% run-time.
runtimes.
We show a proof-of-concept of \clocq's impact on KB-QA 
% we show empirically that QA systems can benefit from \clocq,
by feeding the
% KB subset
output of \clocq into the
% downstream
popular QA system GRAFT-Net~\cite{sun2018open},
and obtain significant boosts in answering performance.
%while keeping \textit{} competitive and the \textit{size of this space} tractable.
% Code for \clocq is available anonymously at \textcolor{red}{\url{clocq.dropbox.com}}.
% All data and results will also be made public upon acceptance.

\myparagraph{Contributions} We make the following salient contributions:
\squishlist
    % this work could (ambitiously) trigger a new task of better contextualization
    % make it clear: answer search
  \item identifying \textit{answer search space reduction} 
    as a 
    %distinct
    critical
    % core
    % KB-QA task;
    task in KB-QA pipelines;
    % with intuitive metrics;
    % that can improve downstream performance of any QA system;
    \item proposing the \clocq % top-$k$ disambiguation
    method
    %that performs this reduction via a
    for % efficient computations of  
    computing 
    answer-containing
    KB subsets with scored lists and the threshold algorithm;
    % subsets of the KB with high answer presence;
    % based on
    % joint
    % a top-$k$ disambiguation technique for question phrases;
    %relying on\textit{ scored lists} and the \textit{threshold algorithm};
    \item conducting extensive experiments that show the superiority of \clocq over a number of baselines using NED; 
    % identifying the trade-off between answer presence and search space size as a major factor for QA systems;
    \item devising a novel KB indexing scheme % based on a fact-centric view of the KB
    that is shown to notably improve runtimes for all methods, including baselines;
    % , and show its positive effect on the runtime of \clocq and baselines;
    \item releasing 
    the complete \clocq code\footnote{\textbf{\url{https://clocq.mpi-inf.mpg.de}}}
    % code link not allowed at CIKM this year
    % (anonymized at \url{https://bit.ly/3bE64ZW})
    % as a module
    along with a \textbf{Web API}
    that any QA system developer can use for
    % KB-QA method can leverage for
    quickly exploring algorithms over much smaller KB subsets.
    % effective KB search space reduction. 
    % , and the effect on the downstream application. 
%    \clocq as a \textit{public interface to KBs} at \textcolor{red}{www.anon.com}, that any QA model can seamlessly operate over, that bypasses the non-trivial resource requirements of processing entire KBs.
    % we should not undersell the importance of this interface, lots of people do not have the memory resources to operate over the whole KB 
\squishend

% !TEX root = ../2021-sigir-sp-clocq.tex
\section{Concepts and Notation}
\label{sec:concepts}

\begin{table} [t]
	\centering
	\caption{Notation for salient concepts in \clocq.}
	\vspace*{-0.3cm}
	\resizebox{\columnwidth}{!}{
		\begin{tabular}{l l}
			\toprule
			\textbf{Notation}				                        & \textbf{Concept}						            \\ \toprule
			$K$	            	                                    & Knowledge base                                    \\
			$x$	            	                                    & KB item                                           \\			
			% $\langle s, p, o, qp_1, qo_1 \dots qo_r \rangle$      & Fact in $K$                                       \\
			$\langle s, p, o, qp_1, qo_1, \dots qo_r \rangle$       & Fact in $K$                                       \\
			$NF(x)$                                                 & 1-hop neighborhood of $x$ (set of facts)          \\
			$NI(x)$                                                 & 1-hop neighbors of $x$ (set of KB items)          \\ \midrule
			$q = \langle q_1 \dots q_{m} \rangle$		            & Question, keywords/phrases in question            \\
			% $\langle x_1 \dots x_{m} \rangle$		                & Combination of KB items mapping to $q$            \\ 
		    %  $l_{i}$		                                        & List of KB items for $q_i$                        \\
			$s$                                                     & Scoring signal                                    \\
			$l_{is}$		                                        & Score-ordered KB-item list for $q_i$ and $s$      \\
 			$\mathcal{S}(q)$                                        & Search space of facts for question $q$            \\ \bottomrule
     	\end{tabular} }
	\label{tab:notation}
	\vspace*{-0.5cm}
\end{table}

% this section and examples extremely important to set the right expectations and reduce confusion

We now introduce % salient
concepts necessary for
% a precise
understanding
% of
\clocq.

\myparagraph{Knowledge base} A knowledge base
$K$
is a compilation of facts. 

\myparagraph{Fact}
A fact is 
a \struct{<subject, predicate, object>} triple, that is optionally augmented by \struct{<qualifier predicate, qualifier object>} pairs which specify context information for the main triple.
For example, \struct{<2018 FIFA World Cup Final, participating team, France national football team; location, Luzhniki Stadium; point in time, 15 July 2018>}
is such a fact, where the first three items constitute the main triple, and the last four make up two qualifier predicate-qualifier object tuples.
Subjects of facts are entities, while objects are entities, types or literals. Predicates denote relationships between the other categories.
% Subjects are entities, while objects are entities (\struct{Paul Pogba}), types (\struct{footballer}) or literals (\struct{15 July 2018}). Predicates (\struct{goal scored by}) denote relationships between the other categories.

% , where each fact is 
% a \struct{<subject, predicate, object>} triple, that is optionally augmented by \struct{<qualifier predicate, qualifier object>} pairs which specify additional information for the main triple (e.g. <\struct{2018 FIFA World Cup Final, participating team, France national football team; no. of goals scored, 4}>). Subjects are entities, while objects are entities (\struct{Paul Pogba}), types (\struct{footballer}) or literals (\struct{15 July 2018}). Predicates (\struct{goal scored by}) denote relationships.
% between the other classes.

% {\color{blue}
\myparagraph{KB item}
% Unlike standard NED which restricts itself to linking individual named entities,
% we also consider types, predicates
% and general concepts (e.g., \struct{football})
% as candidates for disambiguation. 
% as candidates for inclusion in the question-specific search space. 
We refer to entities (\struct{2018 FIFA World Cup Final}), predicates (\struct{participating team}), types (\struct{footballer}) and literals (\struct{15 July 2018})
% all of these
as {\em KB items} $x$.
% I think it's risky to write the sentence below, as most tools don't make a distinction between types and entities, and can thus detect them as well
% Types, in particular, play a very useful role,
% and are disregarded by most NED tools.
% }%\color

\myparagraph{1-hop neighborhood}
This is defined as $NF(x)$ of a KB item $x$ and is given by all facts in which $x$ occurs. The set of KB items $NI(x)$ in the 1-hop neighborhood of $x$ is termed as its \textit{1-hop neighbors}.

% \myparagraph{h-hop neighborhood}
% The \textit{$h$-hop neighborhood} for a specific KB item is then given by all facts of its $(h\text{-}1)$-hop neighbors. The $h$-hop neighbors is the set of distinct items within these facts.

\myparagraph{Question} A question $q$ is specified by a sequence of keywords $\langle q_1, q_2, \ldots q_m \rangle$, where stopwords are not considered.
For our running example \utterance{Who scored in the 2018 final between France and Croatia?}, we would have $q = $ \utterance{<scored, 2018, final, france, croatia>}. 
Without loss of generality, 
$q_i$ may also be a phrase (\phrase{2018 final}).

\myparagraph{Answer} An answer
$\{a\}$
to
% the question
$q$ is a small set of KB entities or literals that satisfy the intent in $q$ (\struct{\{Paul Pogba, Ivan Perisic, \dots \}}).

% \myparagraph{Index list}
\myparagraph{Score-ordered list}
These are
% score-ordered
% ranked 
lists
% $L = \{l\}$
$\{l\}$
that hold KB items $\{x\}$, sorted in descending order of some relevance score. Depending upon the situation, we can have one list $l_i$ per question term $q_i$, or one list $l_{is}$ per score $s$ per $q_i$.

% \myparagraph{Item tuple} A KB-item tuple
% % $t$
% corresponds to a set of KB items $\{x\}_{i=1}^m$, where each $q_i$ maps (like a function) to exactly one $x_i$
% % , and vice versa
% (e.g. \struct{\{goal scored by, 2018 FIFA WC final, France national football team, Croatia football team\}} is an item-tuple for our running example question).

\myparagraph{Search space} A search space $\mathcal{S}(q)$ for a question $q$ is a set of facts $\mathcal{S}(q) \subseteq K$, that is expected to contain each $\{a\}$.
% within the facts necessary to answer $q$, along with other KB facts
For example, $\{\langle$ \struct{2018 FIFA World Cup Final, goal scored by, \textbf{Paul Pogba}; for team, France football team} $\rangle$, $\langle$ \struct{2018 FIFA World Cup Final, goal scored by, \textbf{Ivan Perisic}; for team, Croatia football team} $\rangle, \ldots \}$ comprise a search space for the running example question, where the answers are shown in \textbf{bold}.

% !TEX root = ../2021-sigir-sp-clocq.tex
\section{KB representation and storage}
\label{sec:kb-storage}

\begin{figure} [t]
	\centering
	\includegraphics[width=\columnwidth]{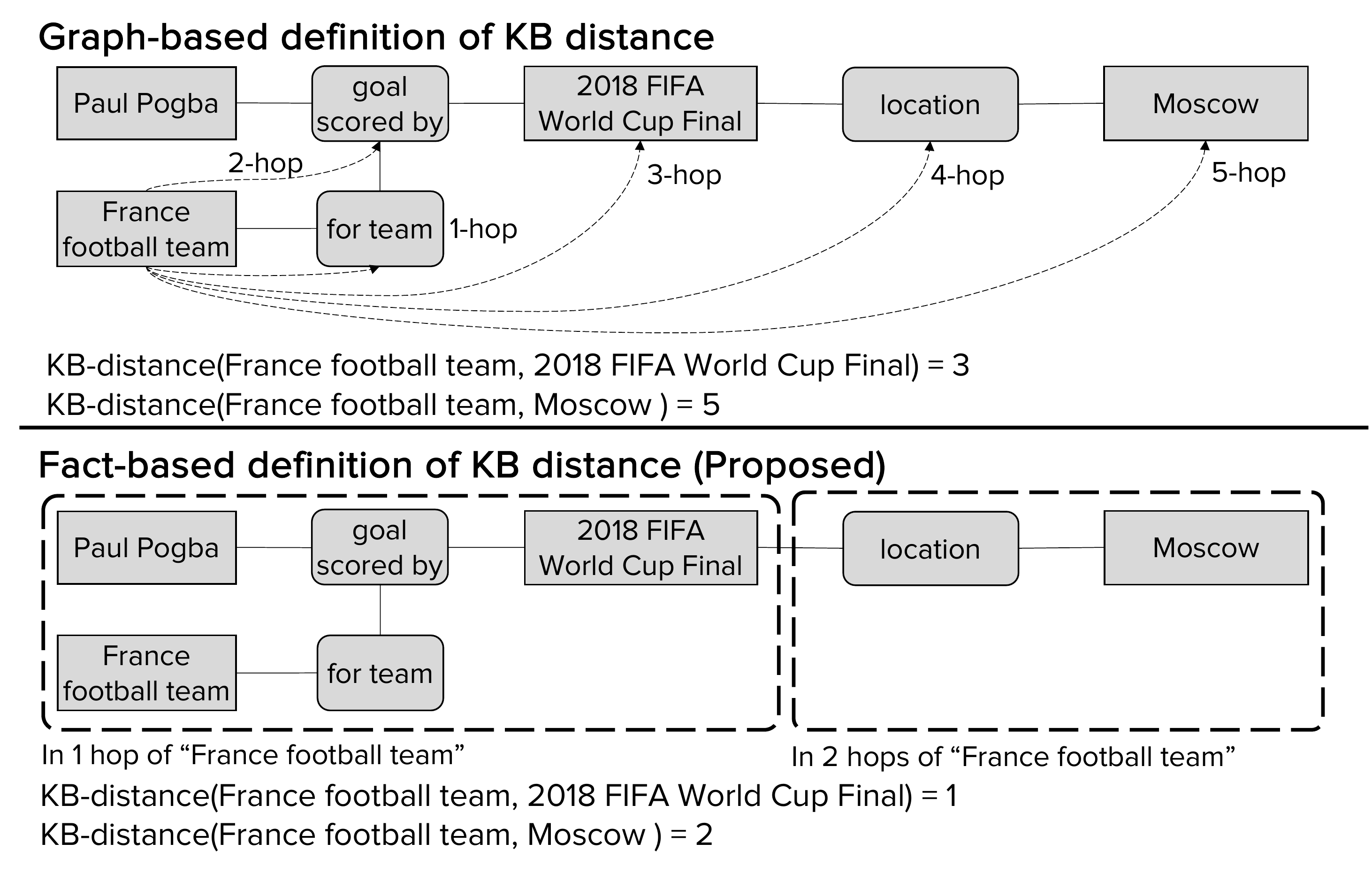}
	\vspace*{-0.7cm}
	\caption{Fact-based definition of KB neighborhoods.} 
	\label{fig:hops}
	\vspace*{-0.5cm}
\end{figure}

% % positioning fig. here for early visibility
\begin{figure*} [t]
	\centering
	\includegraphics[width=\textwidth]{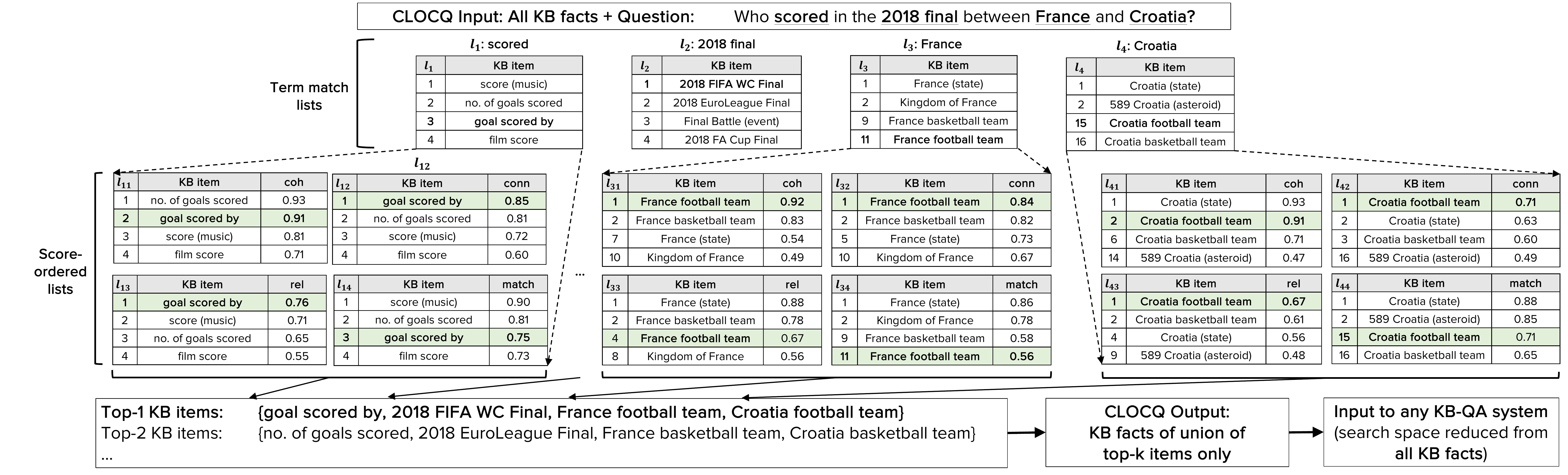}
	\vspace*{-0.7cm}
	\caption{Illustrating the workflow of \clocq for our running example. Input of \clocq: Question + All KB facts; Output of \clocq: Disambiguated question-relevant KB items + Only KB facts with these items.} 
	\label{fig:clocq}
	\vspace*{-0.35cm}
\end{figure*}

% \subsection{Desiderata}
One of the recurrent requirements in QA and specifically in answer search space reduction is to retrieve the facts of a given KB item (like entities returned by an NED system).
Existing KBs are stored as collections of RDF triples.
% , e.g., as NTriples dumps\footnote{\url{https://dumps.wikimedia.org/wikidatawiki/entities/}}
% or in more specialized formats like HDT RDF\footnote{\url{https://www.rdfhdt.org/datasets/}}.
% ~\cite{fernandez2013binary}.
One can query these triple stores in SPARQL-like languages: however, the functionality of fact-retrieval is not built-in, and getting all facts of a single item may often entail issuing a substantial volume of queries (explained later).
% the I/O latency is not enough to guarantee sub-second runtimes for a system like \clocq that postprocesses these retrieval results. 
The consequence is that the total time taken for this step can often be too high, and this is detrimental to any system that relies on these retrieval results. 
As a result, we devise our own KB representation and storage, that are detailed in this section.
% In this section, we
% explain how this was done.
% % introduce
% the KB interface that is used within \clocq. 
% We first motivate the need for a new KB index and introduce the underlying definition of a fact-centric neighborhood. Then we discuss the corresponding KB storage and indexing.
% We conduct experiments comparing our fact-centric KB index with existing KB interfaces in Section~\ref{sec:exp}.

\myparagraph{Concerns with a triple-based KB view}
% \label{subsec:triples}
In standard triple stores, facts containing qualifiers are stored in a \textit{reified} form.
Qualifiers are conceptually modeled as <qualifier predicate, qualifier object> pairs that are appended to the main triple.
However, this is not amenable to store in a uniform triple store. Reification is a technical trick that stitches the main triple with its qualifiers using fact-specific identifiers, also referred to as dummy nodes, and at the same time achieves a ``triplified'' format for all nuggets of information.
For example, the single fact
\struct{<2018 FIFA World Cup Final, participating team, France football team; location, Luzhniki Stadium; point in time, 15 July 2018>}
in reified form would be represented as a set four triples:
\struct{<2018 FIFA World Cup Final, participating team, fact-id>, <fact-id, participating team, France football team>, <fact-id, location, Luzhniki Stadium>, <fact-id, point in time, 15 July 2018>}. 

Joining reified triples into 
their original facts
% a \struct{<s, p, o; qp$_1$, qo$_1$; qp$_2$, qo$_2$>} form
is more amenable to downstream use. % (third row in Table~\ref{tab:notation}).
% An outcome of reification is that triples of the same fact may be \textit{scattered} across the KB dump:
However, such an aggregation
% of all parts of a fact
requires the execution of thousands of structured queries over the KB (equivalently, matching a large number of triple patterns). For example, querying for the triples of \struct{France football team} with this item in the object position will only match the second reified triple above; the whole fact needs to be reconstituted using sequential lookups.
Moreover, this needs to be done for every reified fact that the KB item 
% (\struct{France football team} in this case)
is a part of, which are often several thousands, and additional lookups are also necessary
to get facts 
with the item 
% where this appears 
as a subject.
% and missing parts of the facts (either the main triple or the qualifiers) need to be retrieved and joined
% and needs significant postprocessing time for joining, and bringing to a
% Finally, complex questions contain references to multiple KB items: the process above needs to be repeated for every one of them.
% In total, explaining the high query volume originally indicated.
% Additionally, query-based retrieval needs
% % three distinct
% multiple initial
% lookups: for instance, facts of entities needs consideration of whether the entity appears as a subject, object, or qualifier object.
% This again entails non-trivial overhead. These can of course be combined into one complex query, but this would usually have an even higher runtimes.

\myparagraph{A fact-based view as a solution}
% \label{subsec:soln}
This motivates us to adopt a fact-based view of the KB, that we instantiate as follows.
We start with a standard RDF triple dump. 
% Following existing works, we assign each KB item a unique ID~\cite{fernandez2013binary, urbani2016kognac}. 
We aggregate all reified triples by their fact-id upfront, remove the respective dummy nodes, and postprocess them to the form shown in Table~\ref{tab:notation} (third row).
% corresponding qualifier information to obtain a set of facts, our KB $K$.
Two different indexes are then established: one stores the $1$-hop neighborhood of every item ($x\mapsto NF(x)$), 
and the other stores the set of $1$-hop neighbors of each KB item ($x\mapsto NI(x)$). 
Instead of using alpha-numeric strings that are typical of most raw dumps, KB items are integer-encoded~\cite{fernandez2013binary,urbani2016kognac}.
To reduce the memory footprint, both indexes
% store individual records as start positions and offsets
use appropriate pointers
inside their representations.
% and both indexes are using pointers for facts and KB items,
%%% FURTHER DISCUSSION
The final set of facts obtained this way is referred to as our KB $K$. % in all later sections. % that follow.
% We are now in a position to define the neighborhood of a KB item $x$ under this fact-centric view:
% % \myparagraph{1-hop neighborhood}
% The \textbf{1-hop neighborhood} $NF(x)$ of a KB item $x$ is given by all facts in which $x$ occurs. The set of KB items $NI(x)$ in the 1-hop neighborhood of $x$ is termed as its \textit{1-hop neighbors} (Sec.~\ref{sec:concepts}).

With a fact-based indexing, at runtime, the $1$-hop neighborhood of an item can simply be \textit{looked up}, eliminating the need for expensive querying or joining. Further, the index of $1$-hop neighbors allows for very fast computation of KB distances:
two KB items $x_1, x_2$
% ($|NI(x_2)| \leq |NI(x_1)|$)
are within $1$-hop distance if $x_1 \in NI(x_2)$, and in $2$-hop distance if $NI(x_1) \bigcup NI(x_2) \neq \emptyset$ (via set-overlap tests). This proves decisive
% conducting
for
connectivity checks later on (Sec.~\ref{subsec:itemwise-topk}).

\myparagraph{Additional benefits of a fact-based view}
% \label{subsec:addl}
When a postprocessed fact is directly modeled as a graph (Figure~\ref{fig:hops} top), traditional distance conventions in graphs would imply that even KB items that are part of the same fact may be at a high distance of three (\struct{France football team} and \struct{2018 FIFA World Cup Final}). Distances to KB items in connected facts may be even higher, like five (\struct{France football team} and \struct{Moscow}). 1-hop and 2-hop neighborhoods are vital intuitions of close proximity in KB-QA and such arbitrary distance conventions are far from ideal. In a fact-centric view, \struct{France football team} and \struct{2018 FIFA World Cup Final} are now at a distance of 1, while \struct{France football team} and \struct{Moscow} are 2 hops apart (Figure~\ref{fig:hops} bottom): this is more practical in terms of several KB-related applications. Our approach lifts qualifiers to first-class citizens, this way enhancing the expressiveness of the QA method within limited neighborhoods.

The concept of a KB neighborhood in the literature is primarily entity-centric. An ideal representation should enable definitions that uniformly apply to entities, predicates, types and literals. Predicates are often modeled as edge labels, and this precludes a seamless notion of neighborhood. A fact-based neighborhood can easily be envisioned for all types of KB items. 

\section{The \clocq Method}
\label{sec:method}

% We first briefly explain how \clocq stores and indexes the KB, and then discuss 
% In this section,
We now explain
the complete \clocq workflow
% as shown in
(illustrated in Fig.~\ref{fig:clocq}).

\subsection{Retrieving candidate KB items per term}
\label{subsec:kb-model}

% \myparagraph{Storing the KB} It is of utmost importance that \clocq completes its operation with minimal runtime: this necessitates additional measures for efficiency. First, we assign an integer ID to each unique KB-item (entities, predicates, types and literals) to bring down memory requirements from their original string representations. Qualifiers are usually stored in KB dumps via \textit{reification}~\cite{hernandez2015reifying,suchanek2007yago} that introduces dummy nodes to store a fact with (possibly multiple) qualifiers in a fragmented but ``triplified'' form. This results in significant joining effort at runtime to assemble the pieces of a single fact: \clocq does this joining upfront,
% as 
% % An example is
% shown in the example fact in our definitions (Sec.~\ref{sec:concepts}). Finally, \clocq indexes facts of every KG-item (similar to inverted indexes for term-document lists) that speeds up lookups, and precomputes item neighbors in the
% % equivalent graph representation of the KB
% KB graph
% to expedite connectivity checks (via set-overlap tests).

% \subsection{Creating initial lists}
% \label{subsec:naive-topk}

\myparagraph{Creating term match lists}
% A disambiguation process over a sentence-like text input in IR/NLP usually works by: i) quickly creating candidate lists per keyword, and subsequently, ii) using smart signals to locate the best set of candidates that
% jointly
% optimizes some criterion.
Consider our running example question: \utterance{Who scored in the 2018 final between France and Croatia?}
As our goal is to disambiguate keywords or phrases in the question (\phrase{scored}, \phrase{2018 final}, \phrase{France}, \phrase{Croatia}) to items in a KB,
% and are given a sequence of question words or phrases,
we first % take the natural step of
collect candidates from the KB using a standard lexical matching score (like TF-IDF or BM25) for each question term $q_1 \ldots q_m$ ($m = 4$ in our example, stopwords are dropped).
% elastic search with Wikidata API only in experiments... + item labels and aliases
Here $q_i$ is analogous to a search query, while each item $x$ in the KB resembles a document in a corpus. This ``document'' is created by concatenating the item label with textual aliases and descriptions available in most KBs~\cite{bollacker2008freebase,vrandevcic2014wikidata}.
This results in $m$ ranked lists $\langle l_1 = \{x_{11}, x_{12}, \ldots \}; l_2 = \{x_{21}, x_{22}, \ldots \}; \ldots l_m = \{x_{m1}, x_{m2}, \ldots \} \rangle$ of KB items $x_{ij}$, one list $l_i$ for each $q_i$, scored by degree of match between question tokens and KB items.
% (top half in Fig.~\ref{fig:clocq}).
A ranked lexical match list (ideal disambiguation in \textbf{bold}) for \phrase{scored} could look like: $l_1=$\struct{$\langle$1: score (music), 2: no. of goals scored, \textbf{3: goal scored by}, 4: film score, ...$\rangle$}, while that for \phrase{Croatia} could be: $l_4=$\struct{$\langle$1: Croatia (state), 2: 589 Croatia (asteroid), ..., 15: \textbf{Croatia football team}, ..., 19: Croatia basketball team, ...$\rangle$}.
Note that the best matching KB item $x_i^*$ for $q_i$ can sometimes be \textit{very deep in individual lists} $l_i$ (\struct{Croatia football team} is at rank 15 in $l_4$).
% Thus, simply extending top-1 NED systems to return top-5 disambiguations or so would not suffice.

% Now that we have ranked lists $l_1 \ldots l_m$ of KB-items per query term, a standard way in IR of finding the best candidates would be to apply rank aggregation~\cite{farah2007outranking} on these individual lists. % This phase is roughly similar to pseudorelevant candidate document retrieval for a keyword query under a disjunctive model, the main difference being that here we have $m$ lists for an $m$-keyword query, whereas in the usual IR scenario we have only one list for the entire query.
% However, note that each of these $m$ lists potentially contains a very large number of KB-items $x_{ij}$. Thus, typical rank aggregation methods would be very expensive as they require scoring all candidate item-tuples from these lists, via an exhaustive enumeration of the Cartesian product of these lists ($|l_1| \times |l_2| \times \ldots |l_m|$).
% % These lists can be considered analogous to rank-join lists where each list potentially holds a different set of items.

Next, each list $l_i$ is traversed up to a depth $d$ to fetch the top-$d$ items (computational cost $O(m \cdot d)$), that are per-term question-relevant KB candidates for the next phase of \clocq. The goal is to find
% \textit{cross combinations}
\textit{combinations}
of KB items
% KB item-tuples
$\langle x_i \rangle_{i = 1}^m$ that best match the question,
since these items have a high likelihood of having the \textit{answer within their facts} $\bigcup_{i = 1}^m NF(x_i)$.
For instance, an ideal combination for us would be:
\struct{\{goal scored by, 2018 FIFA WC final, France football team, Croatia football team\}}.
These
% tuples
combinations
come from the Cartesian product of items in the $m$ lists, and would have $d^m$ possibilities if each
% tuple
combination
is explicitly enumerated and scored. This is cost-prohibitive: since we are only interested in some top-$k$
% tuples,
combinations,
as opposed to a full or even extended partial ordering,
% of these tuples,
a more efficient way of doing this would be to apply top-$k$
% algorithms~\cite{mackenzie2020examining,anh2006pruned,ilyas2008survey}
algorithms~\cite{mackenzie2020examining,anh2006pruned,ilyas2008survey}.
% on these ranked lists.
These prevent complete scans and return the top-$k$ best
% tuples
combinations
efficiently.

\subsection{Computing relevance signals for each item}
\label{subsec:itemwise-topk}
% calling these relevance scores is good for the IR community

% Thus, we propose a smart way of using top-$k$ algorithms to overcome this challenge.
To go beyond shallow lexical matching, our proposal is to
% rather
construct \textit{multiple lists per question token, each reflecting a different relevance signal}, and to apply top-$k$ algorithms \textit{on these lists} to obtain the disambiguation of each question token individually.
% quick justification on our indicators
Unlike prior works on NED that are restricted to individual named entities~\cite{ferragina2010tagme,hoffart2011robust,li2020efficient},
\clocq includes mentions of types, predicates and general concepts in the input question
and maps them to KB items.
A candidate
% tuple
KB item combination
that fits well with the intent in the question is expected to have high semantic coherence and high graph connectivity (these can be viewed as proximity in latent and symbolic spaces) within its constituents, as well as match the question well at global and term-levels.
These motivate our four indicators of relevance for each item $x_{ij}$ in list $l_i$ below
(the cost of this scoring is $O(m^2 \cdot d^2)$: while this looks expensive, it is still fast with a parallelized implementation).
% , as demonstrated in our experiments).
% semantic coherence, knowledge graph
% % density,
% connectivity, question-level semantic relatedness, and term-level lexical match:
% \squishlist

\myparagraph{Coherence}
\clocq targets a joint disambiguation of question-relevant KB items.
It thus considers semantic coherence and graph connectivity, which are inherently defined for KB item pairs, instead of single items.
Therefore, we need a technique to convert these signals into item-level scores.
% \item \textbf{Coherence:}
% Since \clocq targets
% a joint
% disambiguation of question-relevant 
% % combination
% KB items, it considers two
% % itemset-level
% % context-level
% signals that are defined only for combinations of KB items.
% Nevertheless,
% % item-level lists $l_{ij}$ are required,
% we need to construct item-level lists $l_{ij}$, and
% so we need a technique to convert such signals into item-level scores.
% The first signal is the
The first signal, semantic coherence, % between a
% tuple
% combination
% of KB items, that
is transformed to an item-level score using the max operator.
More precisely, the coherence score of an item $x_{ij}$ is defined in Eq.~\ref{eq:coh} as the maximum item-item similarity (averaged over pairs of lists) this item can contribute to the
% item-tuple
combination,
where pairwise similarity is obtained by the cosine value between the embedding vectors of two KB items (min-max normalized from $[-1, +1]$ to $[0, 1]$):
% wikipedia2vec in experiments
    \vspace*{-0.1cm}
\begin{equation}
    coh(x_{ij}) = \frac{1}{m - 1} \sum_{k \neq i} \max_l cosine(\Vec{x_{ij}}, \Vec{x_{kl}})
    \vspace*{-0.1cm}
    \label{eq:coh}
\end{equation}
%\item \textbf{Connectivity:}

\myparagraph{Connectivity}
This is the second
% tuple-level
context-level
signal
% used
in \clocq, and captures a very different form of proximity. 
% Every KB can be viewed as an equivalent knowledge graph (KG), where entities, predicates and other KB items are nodes, and edges run between components of the same fact~\cite{lu2019answering,christmann2019look}. 
% We define KB items that are \textit{part of the same fact} to be in the 1-hop neighborhood of each other, those that are connected \textit{via members of another fact} as in the 2-hop, and so on. 
% As an example, all KB items (including predicates) in <\struct{2018 FIFA World Cup Final, participating team, France football team; goal scored by, Paul Pogba}> are in 1-hop of each other, and are $2$-hops away from \struct{Lagny-sur-Marne}, due to the connector fact \struct{<Paul Pogba, place of birth, Lagny-sur-Marne>}. 
% KB distances are defined % as outlined
% in Section~\ref{subsec:addl}.
We assign items in $1$-hop of each other to have a distance of $1$ (recall KB-distance computations
% and applied
from Sec.~\ref{sec:kb-storage}), those in 2-hops to have a distance of $2$, and $\infty$ otherwise (most KB items are in 3 hops of each other, and thus distance > 2 hops ceases to be a discriminating factor). We define connectivity scores as the inverse of this KB distance, thereby obtaining $1, 0.5$, and $0$, respectively for $1$-, $2$-, and $>2$-hop neighbors. Connectivity as a
% tuple-level
context-level
signal is converted to an item-level
% \textit{density} (= pairwise connectivity)
score analogously using max aggregation over pairs.
% We empirically observed averaging (instead of max) to be biased towards frequent KB-items.
We thus define
connectivity
% density
($\in [0,1]$) of $x_{ij}$ in Eq.~\ref{eq:conn}:
\vspace*{-0.1cm}
\begin{equation}
    % dens(x_{ij}) = \frac{1}{m - 1} \sum_{k \neq i} \max_l conn(x_{ij}, x_{kl})
    conn(x_{ij}) = \frac{1}{m - 1} \sum_{k \neq i} \max_l conn(x_{ij}, x_{kl})
    \vspace*{-0.1cm}
    \label{eq:conn}
\end{equation}

% \item \textbf{Question relatedness:}
\myparagraph{Question relatedness}
We estimate semantic relatedness of the KB item $x_{ij}$ to the overall input question $q$ by averaging pairwise cosine similarities (same min-max normalization as for coherence) between the embeddings of the item and each term $q_i$ in Eq.~\ref{eq:qmatch}. To avoid confounding this estimate with the question term for which $x_{ij}$ was retrieved, we exclude this from the average to define semantic relatedness as:
\vspace*{-0.1cm}
\begin{equation}
    rel(x_{ij}) = \textrm{avg}_{q_i \neq q_k} cosine(\Vec{x_{ij}}, \Vec{q_k})
    \vspace*{-0.1cm}
    \label{eq:qmatch}
\end{equation}

% \item \textbf{Term match:}
\myparagraph{Term match}
This score is intended to take into account the original degree of lexical term match (via TF-IDF, BM25, or similar) for which $x_{ij}$ was admitted into $l_i$.
% While the initial score is computed as the TF-IDF (or similar) weight that $x_{ij}$ received when the KB item list was queried with the term $q_i$,
However, such TF-IDF-like weights are often unbounded and may have a disproportionate influence when aggregated with the other signals, that all $\in [0, 1]$. Thus, we simply take the reciprocal rank of $x_{ij}$ in $l_i$ as the
% term
match score (Eq.~\ref{eq:tmatch}) to have it in the same $[0, 1]$ interval: 
\vspace*{-0.1cm}
\begin{equation}
    match(x_{ij}) = 1/rank(x_{ij}, l_i)
    \vspace*{-0.1cm}
    \label{eq:tmatch}
\end{equation}
% \squishend
% Here $q_i$ is analogous to a query, while each item $x$ in the KB resembles a document in a corpus. This ``document'' is created by concatenating the item label with textual aliases and descriptions available in most KBs.

\subsection{Finding top-$k$ across sorted lists}
\label{subsec:ta}

We now sort each of these $4 \cdot m$ lists in descending score-order. Note that for each $q_i$, all lists $l_{is}$ hold the same items (those in the original $l_i$). Fig.~\ref{fig:clocq} shows lists $l_{is}$ in the center. Top-$k$ algorithms operating over such multiple score-ordered lists, where each list holds the same set of items, require a monotonic aggregation function over the item scores in each list~\cite{anh2006pruned,DBLP:conf/vldb/BastMSTW06,buckley1985optimization,fagin2003optimal}. Here, we use a linear combination of the four relevance scores as this aggregate: $aggScore(x_{ij}) = h_{coh} \cdot coh(x_{ij}) + h_{conn} \cdot conn(x_{ij}) + h_{rel} \cdot rel(x_{ij}) + h_{match} \cdot match(x_{ij})$, where hyperparameters are tuned on a dev set, and $h_{coh} + h_{conn} + h_{rel} + h_{match} = 1$.
Since each score lies in $[0, 1]$, we also have $aggScore(\cdot) \in [0, 1]$. 

\myparagraph{Threshold algorithm} 
We use the 
%well-known Fagin's 
%%%GW: don't emphasize "Fagin's""; IR folks believe they invented all this
threshold algorithm (TA)
% (see~\cite{buckley1985optimization} for a similar method)
over these score-ordered lists with early pruning~\cite{fagin2003optimal}.
% to instantiate the family of top-$k$ operators.
TA is run over each set of $4$ sorted lists $\langle l_{i1}, l_{i2}, l_{i3}, l_{i4} \rangle$, corresponding to one question term $q_i$, to obtain the top-$k$ best KB items $\{x_i^*\}_k$ per $q_i$, as follows:
we perform a sorted access (SA) in parallel on each of the four sorted lists for each $q_i$. For each item $x_{ij}$ seen with SA, we fetch all its
% values
scores
$coh(x_{ij}), conn(x_{ij}), rel(x_{ij})$ and $match(x_{ij})$ by random access (RA). We compute $aggScore(x_{ij})$, and if $x_{ij}$ is one of the top-$k$ scoring items so far, we remember this. For each list $l_{is}$, let $\hat{l_{is}}$ be the score of the last item seen under SA.
Given that lists $l_{is}$ are sorted, this score $\hat{l_{is}}$ is the maximum value that could be observed in the unknown part of the list.
We define the \textit{threshold} $\delta$ as the aggregate of these maximum scores, i.e. 
% $\delta = \sum_s h_s \cdot \hat{l_{is}}$. 
$\delta = h_{coh} \cdot \hat{l_{i1}} + h_{conn} \cdot \hat{l_{i2}} + h_{rel} \cdot \hat{l_{i3}} + h_{match} \cdot \hat{l_{i4}}$.
When $k$ items have been seen whose aggregate is at least $\delta$, TA is terminated and the top-$k$ KB items are returned.
% We perform a sorted access (SA) in parallel on each of the four sorted lists for each $q_i$. For each item $x_{ij}$ seen with SA, we fetch all its values $coh(x_{ij}), conn(x_{ij}), rel(x_{ij})$ and $match(x_{ij})$ by random access (RA). We compute $aggScore(x_{ij})$, and if $x_{ij}$ is one of the top-$k$ scoring items so far, we remember this. For the $j^{th}$ item in each list $l_{ij}$, let $\hat{l_{ij}}$ be the score of the last item seen under SA. We define the \textit{threshold} $\delta$ as the aggregate over the items $\hat{x_{ij}}$ from each list $j$. When $k$ items have been seen whose aggregate is at least $\delta$, TA is terminated and the top-$k$ KB items are returned.
Once we have these items $\{x_i^*\}_k$, we take the union 
% of these items 
$\bigcup_{i = 1 \ldots m}\{x_i^*\}_k$ to create our final combination of KB items. \textit{KB facts} of items in this final list comprise $\bigcup_{i = 1 \ldots m}\{NF(x_i^*)\}_k$,
% not necessary
% (often referred to as 1-hop neighborhood)
which is the output search space $\mathcal{S}$ of \clocq and would be passed on to the downstream QA system.

\subsection{Pruning facts of highly frequent KB items}
\label{subsec:ta}

\begin{figure} [t]
	\centering
	\includegraphics[width=\columnwidth]{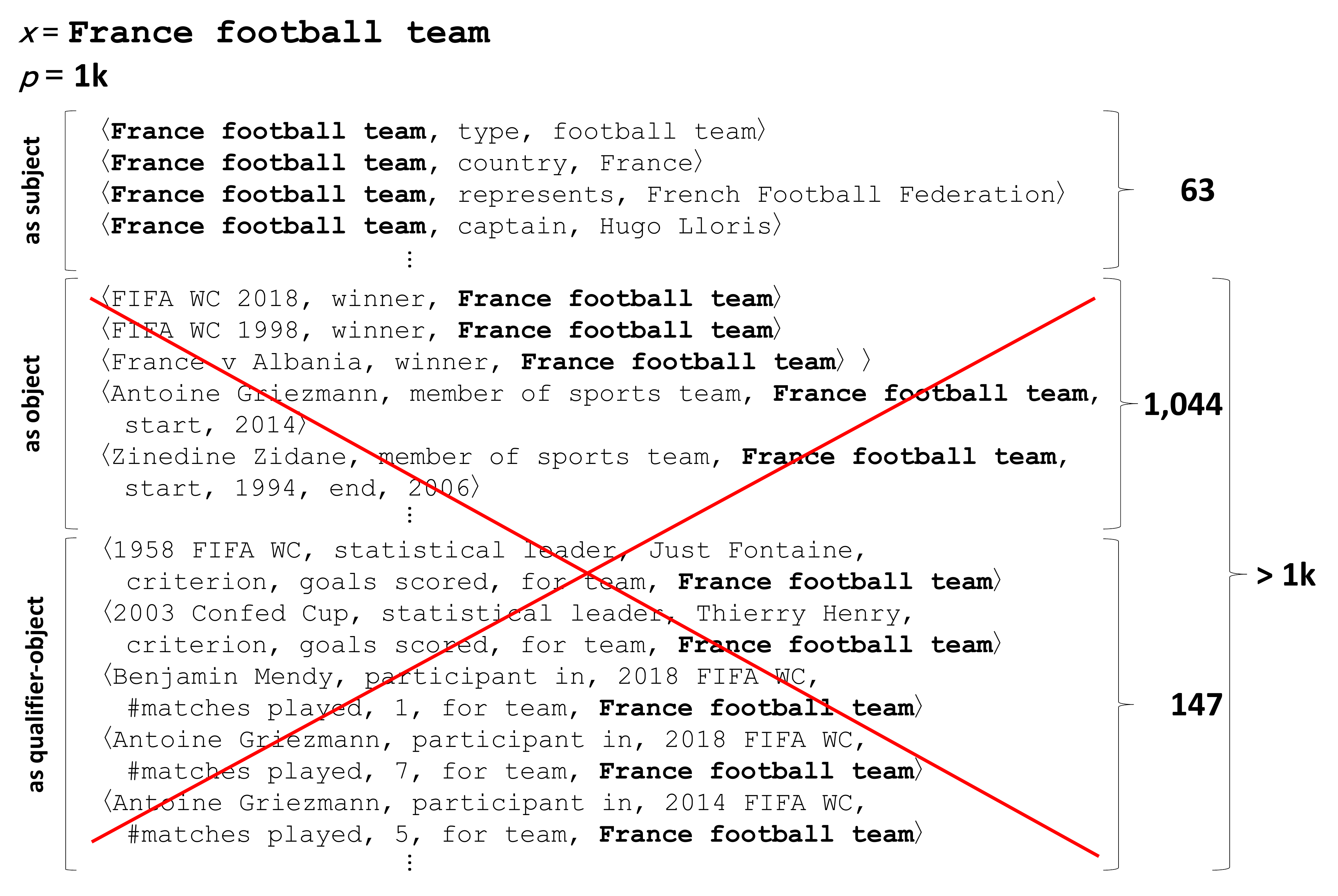}
	\vspace*{-0.7cm}
	\caption{Pruning the search space with parameter $p$.} 
	\label{fig:param-p}
% 	\vspace*{-0.5cm}
\end{figure}

To avoid including all facts of extremely frequent KB items into our search space $\mathcal{S}$ (\struct{U.K.} brings in millions of entities), we use a pruning threshold $p$ % set
as follows. An entity $x$ can appear in a fact as the subject, object or qualifier object, where usually the first role is the most salient. Whenever the last two total more than $p$, we take only the subject facts of $x$ (and all facts otherwise): this is a proxy for keeping only salient facts in
% search space.
$\mathcal{S}$.
% When $x$ is a predicate, 
For disambiguated predicates $x$,
$p$ directly acts as a frequency threshold.
Thus, parameter $p$ essentially controls the amount of potentially noisier facts that goes into $\mathcal{S}$.
Fig.~\ref{fig:param-p} illustrates how the parameter $p$ helps to prune the search space for \struct{France football team} ($p$ is set to $1$k).

\begin{table*} [t] \small
    \caption{Performance of \clocq
	% (default, $k$=1 and $k$=5)
	w.r.t. baselines. Statistical significance of \clocq's answer presence over \textsc{Tagme} and \textsc{Elq}, the strongest baselines, is marked with $\dagger$ and * respectively (McNemar's test as answer presence is a binary variable, with $\mathrm{p} < 0.05$).}
	\vspace*{-0.3cm}
    \newcolumntype{G}{>{\columncolor [gray] {0.90}}c}
	% \resizebox{\textwidth}{!}{
    	\begin{tabular}{l G G G c c c}
    	    \toprule
    	    % \textbf{Benchmark}  & \multicolumn{3}{G}{\textbf{LC-QuAD2.0-CQ~\cite{dubey2019lc} (10k questions)}}     & \multicolumn{3}{c}{\textbf{ConvQuestions-FQ~\cite{christmann2019look} (1.5k questions)}}  \\
    	    \textbf{Benchmark}  & \multicolumn{3}{G}{\textbf{LC-QuAD2.0-CQ~\cite{dubey2019lc}}}     & \multicolumn{3}{c}{\textbf{ConvQuestions-FQ~\cite{christmann2019look}}}  \\
    	   % \cmidrule(l){1} \cmidrule(l){2-4} \cmidrule(l){5-7}
    	    \midrule
    	     \textbf{Metric $\rightarrow$}  & \textbf{Answer presence}  & \textbf{Search space size}     & \textbf{Runtime}   
    	                                    & \textbf{Answer presence}  & \textbf{Search space size}     & \textbf{Runtime}      \\   
    	     \textbf{Method $\downarrow$}	& \textbf{(Percentage)}     & \textbf{(No. of KB items)}   & \textbf{(Seconds)}   
    	                                    & \textbf{(Percentage)}     & \textbf{(No. of KB items)}   & \textbf{(Seconds)}    \\    \toprule

    	     \textbf{\textsc{Tagme}~\cite{ferragina2010tagme}+\textsc{Hdt}~\cite{fernandez2013binary}} & $76.8$ & $2.9$k &		  	              $1.14$ % LCQuAD 2.0
    	                                               & $69.1$ &                         $1.8$k &		                               $1.43$     \\ % ConvQuestions

    	     \textbf{\textsc{Aida}~\cite{hoffart2011robust}+\textsc{Hdt}~\cite{fernandez2013binary}}	  &		  $60.5$ &                         $2.2$k &		 $0.75$  % LCQuAD 2.0 
    	     &		  $44.4$ &                         $2.2$k &		                              $1.19$ \\    % ConvQuestions
    	     \textbf{\textsc{Earl}~\cite{dubey2018earl}+\textsc{Hdt}~\cite{fernandez2013binary}} ($k$=1)   &		  $53.8$ &                         $1.1$k &		             $2.50$ % LCQuAD 2.0
    	     &		  $44.6$ &                         $1.1$k &		                       $2.49$    \\ % ConvQuestions
    	     \textbf{\textsc{Earl}~\cite{dubey2018earl}+\textsc{Hdt}~\cite{fernandez2013binary}} ($k$=5)    &		  $65.9$ &                         $2.2$k &		             $2.50$ % LCQuAD 2.0
    	     &		  $53.4$ &                         $2.0$k &		                       $2.49$    \\ % ConvQuestions
    	     \textbf{\textsc{Rel}~\cite{van2020rel}+\textsc{Hdt}~\cite{fernandez2013binary}}	        &		  $55.8$ &                          $0.7$k  &		      $0.72$ % LCQuAD 2.0    
    	        &		  $45.6$ &                          $0.4$k  &		               $0.61$    \\ % ConvQuestions
    	     % use only one ELQ for space
    	     \textbf{\textsc{Elq}~\cite{li2020efficient}+\textsc{Hdt}~\cite{fernandez2013binary}}  &		 76.7 &                         $1.1$k &		          $0.62$ % LCQuAD 2.0
    	     & 77.5 &                           $0.6$k 	             &  0.47     \\ % ConvQuestions
    	    \midrule                                                
    	     % \textbf{\clocq: Naive top-$1$}	                 & 76.2             &                         $4.2k$ &		                  $0.50$  % LCQuAD 2.0
    	   %                                                  &		 72.9 &                         $2.8k$ &		                  $0.44$    \\% ConvQuestions
    	    \textbf{$\clocq$} (Default: $k$=Auto, $p$=1k) & $\boldsymbol{82.6}^\dagger$* &    $1.5$k &		  $0.50$   % LCQuAD 2.0
    	                    & $\boldsymbol{84.7}^\dagger$* &     $1.3$k &		  $0.42$  \\ % ConvQuestions     
    	                    \midrule
    	    \textbf{$\clocq$}  ($k$=1, $p$=10k)  & $\boldsymbol{80.0}^\dagger$* &    $3.9$k &		  $0.48$   % LCQuAD 2.0
    	                                & $\boldsymbol{78.4}^\dagger\phantom{*}$ &     $2.3$k &		  $0.39$  \\% ConvQuestions
    	   % \textbf{\clocq (k=5, p=1$k$)}  & $\boldsymbol{85.2}^\dagger$* &    $3.0k$ &		  $0.56$   % LCQuAD 2.0
    	                   % & $\boldsymbol{86.8}^\dagger$* &     $2.7k$ &		  $0.49$  \\% ConvQuestions
    	   % \textbf{$\clocq_3$} (k=3, p=1$k$)  & $\boldsymbol{83.2}^\dagger$* &    $1.8k$ &		  $0.44$   % LCQuAD 2.0
    	   %                 & $\boldsymbol{85.4}^\dagger$* &     $1.5k$ &		  $0.39$  \\% ConvQuestions        
    	    \textbf{$\clocq$} ($k$=5, $p$=100)  & $\boldsymbol{80.9}^\dagger$* &    $0.6$k &		  $0.49$   % LCQuAD 2.0
    	                    & $\boldsymbol{84.2}^\dagger$* &     $0.6$k &		  $0.40$  \\% ConvQuestions        
    	       
    	    \bottomrule
    	\end{tabular} %}
	% \raggedright The highest value in a column is marked in \textbf{bold}.
	\label{tab:main-res}
	\vspace*{-0.3cm}
\end{table*}

\begin{figure} [t]
	\centering
	\includegraphics[width=\columnwidth]{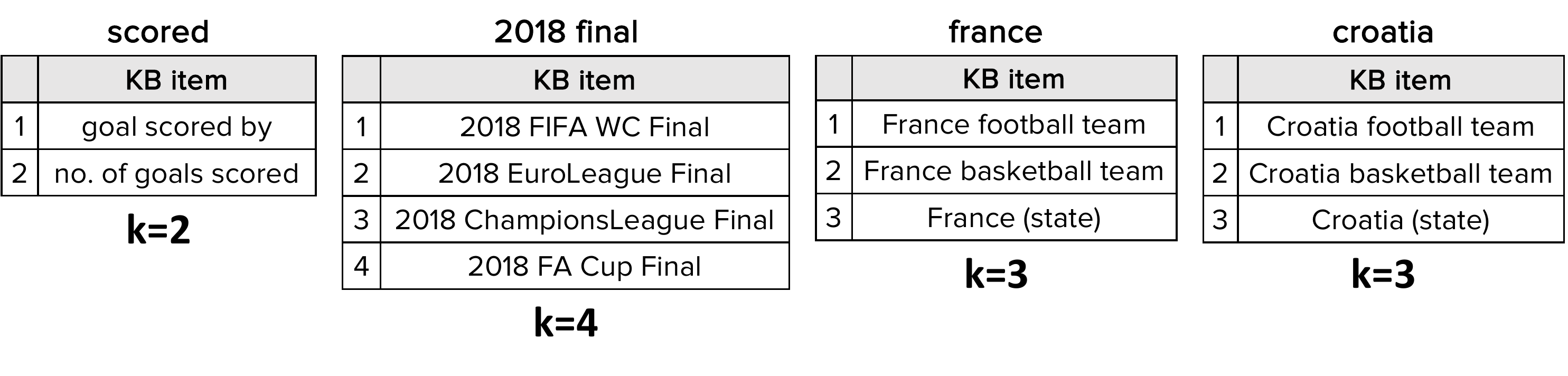}
	\vspace*{-0.7cm}
	\caption{Auto-$k$ setting for running example.} 
	\label{fig:auto-k}
	\vspace*{-0.5cm}
\end{figure}

\subsection{Automatic setting of $k$ and $p$}
\label{subsec:dyn}

The choice of $k$ and $p$ might not always be obvious, and in the methodology described above, it is set globally to the same value for all question words. Therefore, we propose a simple but effective mechanism to automatically choose $k$ and $p$, dynamically depending upon the specific question word. %, as an extension.

\myparagraph{Choosing $\boldsymbol{k}$} Intuitively, one would like to increase $k$ for \textit{ambiguous} question words.
% , if it is inherently . 
For example, \phrase{France} can refer to many KB items. By increasing $k$ one can account for potential disambiguation errors. On the other hand, \phrase{Paul Pogba} is not as ambiguous and hence $k\text{=}1$ should suffice.
The ambiguity of a question word is closely connected to that of uncertainty or randomness. 
The more uncertainty there is in predicting what a word refers to, the more ambiguous it is. This makes \textit{entropy}
% ~\cite{shannon2001mathematical}
a suitable measure of ambiguity.
More specifically, each question word is linked to $d$ KB items. These items form the sample space of size $d$ for the probability distribution. The numbers of KB facts of these items form a frequency distribution that can be normalized to obtain the required probability distribution. We compute the entropy of this probability distribution as the ambiguity score of a question word, and denote it as $ent(q_i)$.
Incidentally, by definition, $0 \leq ent(q_i) \leq \log_2 d$. Practical choices of $k$ and $d$ does not exceed $5$ and $50$ respectively, and hence $k$ and $\log_2 d$ are in the same ballpark ($\log_2 50\text{=}5.6$). This motivates us to make the simple choice of directly setting $k$ as $ent(q_i)$.
Specifically, we use $k = \floor{ent(q_i)} + 1$ to avoid the situation of $k\text{=}0$. % , such that $k \in [1,\floor{\log_2(d)}+1]$.
Fig.~\ref{fig:auto-k} shows a possible ``Auto-$k$'' (automatic choice of $k$) setting for our running example. \phrase{2018 final} is highly ambiguous, and thus $k$ is set to a relatively high value. 
\phrase{France} and \phrase{Croatia} can also refer to various different concepts.
The word \phrase{scored} is relatively unambiguous.

\myparagraph{Choosing $\boldsymbol{p}$} We identify a logical connection between $k$ and $p$: the less uncertainty there is in the disambiguation of a question word (i.e. the lower the $k$), the more facts one wants to include in
$\mathcal{S}$ for this word.
% TODO: add specific example of f here (10^(5-k)), refer to variations (section 6.2)
% k = 1-5 -> p in [...]
% the search space.
On the contrary, for highly ambiguous question words, less facts should be admitted for avoiding a higher amount of noise.
Therefore, we set $p$ automatically,
% based on $k$.
by having $p\text{=}f(k)$.
For example, we could set $p$=$10^{(5-k)}$, such that $p$ is set to a high value ($p$=$10^4$) for $k$=$1$, but for a highly ambiguous word for which $k$=$5$, only subject facts are considered ($p$=$1$).
We experiment with different variations of the function $f$ that meet the desired criterion above. % (Sec.~\ref{subsec:analysis}).

\section{Experimental setup}
\label{sec:exp-setup}

\myparagraph{Benchmarks} We use
% perform experiments on
two recent QA benchmarks:
% of complex questions:
LC-QuAD 2.0~\cite{dubey2019lc} and ConvQuestions~\cite{christmann2019look}.  
%To be exact, 
To make our case, we sampled
$10$k of the more \underline{c}omplex \underline{q}uestions 
%were sampled 
from LC-QuAD 2.0
% more details not necessary for space
(LC-QuAD2.0-CQ in Table~\ref{tab:main-res} with $2$k dev, $8$k test; no training required in \clocq). Complexity is loosely identified by the presence of multiple entities, as detected with Tagme~\cite{ferragina2010tagme}, and/or predicates where main verbs were used as a proxy, detected with Stanza~\cite{qi2020stanza}.
% We make this subset available upon acceptance.
% (henceforth referred to as LC-QuAD2.0-CQ or just LC-QuAD)
% Next, for
ConvQuestions was built for incomplete utterances in conversational QA, but also has 
well-formed complete questions that exhibit
several complex phenomena.
For
ConvQuestions, we considered \underline{f}ull \underline{q}uestions from the benchmark
(ConvQuestions-FQ in Table~\ref{tab:main-res}; $338$ dev, $1231$ test).
% henceforth referred to as ConvQuestions-FQ (this
%many of which are known to be complex~\cite{christmann2019look}.
% There are $1569$ questions, out of which $338$ and $1231$ were used for test and dev, respectively (same $\simeq 1:4$ ratio as in LC-QuAD). 

\myparagraph{Metrics} We use three metrics: i) 
{\em answer presence}, the percentage of times the correct answer is found in the reduced search space; ii) {\em size of the search space} $|\mathcal{S}|$, measured by the number of entities and literals, that would be answer candidates to be considered by the downstream QA engine; 
and iii) {\em runtime}, 
%that is 
% time taken for the method to fetch the desired KB facts, and is measured in seconds.
% end-to-end
%aggregated 
summed over all steps 
% totaled
% over all 
%answering-time 
% question-time
% question-time is confusing
that happen at answering time and measured in seconds.

\myparagraph{Baselines} We compare \clocq with a variety of NED baselines~\cite{ferragina2010tagme,hoffart2011robust,li2020efficient,dubey2018earl,van2020rel}.
% , ranging from 
%the most 
% popular~\cite{ferragina2010tagme,hoffart2011robust} to %the most 
% recent~\cite{li2020efficient,dubey2018earl,van2020rel}.
% (Table~\ref{tab:main-res}).
To provide baselines with competitive advantage w.r.t. efficient retrieval, we use the state-of-the-art HDT RDF~\cite{fernandez2013binary} for KB storage and indexing. 
An example baseline would be \textsc{Tagme+Hdt}. For convenience, we omit the \textsc{Hdt} when referring to baselines in text.
NED systems that link to Wikipedia are mapped to Wikidata using
Wikipedia
% article
URLs that are also present in Wikidata.
Baselines are either run on our data with original code when available, or through APIs.
% Whenever possible,
Internal confidence thresholds were set to zero (no cut-off) in configurable baselines like \textsc{Tagme} and \textsc{Aida}
to allow for as many disambiguations (linkings) as possible, to help boost answer presence. Otherwise, default configurations were retained.
% Naive top-$k$ operating directly on the index-lists is our configurational baseline.

% \myparagraph{Initialization}
\myparagraph{KB cleaning}
We perform all experiments over Wikidata.
% a very popular KB today. 
%specifically, 
The original Wikidata dump contains a large set of facts that are not needed for common question answering use cases. For example, Wikidata contains labels and descriptions in several languages, it provides meta-information for internal concepts (e.g. for predicates), and references for facts, URLs, images, or geographical coordinates. Furthermore, identifiers for external sites such as Spotify ID, IMDb ID, or Facebook ID are stored.
As an initial effort, we pruned all facts containing such information from an N-triples Wikidata dump downloaded on 24 April 2020, such that the size on disk decreased from $1,990$ GB to $450$ GB\footnote{\url{https://github.com/PhilippChr/wikidata-core-for-QA}}.

\myparagraph{Initialization}
After applying our KB index (Sec.~\ref{sec:kb-storage}), the size decreased to $18$ GB on disk.
% we use the 
%$\simeq 2$ 
% 2 TB uncompressed NTriples dump from 26 April 2020 with %$\simeq 12B$ 
% ca. 12B
% triples, on which \clocq 
%KB processing 
% is applied.
Note that we applied the same pruning strategy and underlying Wikidata dump when using \textsc{Hdt} for retrieval, 
% i.e. $NF(x)$ given some specific $x$ and $p$ is exactly the same for the \clocq KB interface and \textsc{Hdt}.
i.e. $NF(x)$ is exactly the same for the \clocq KB interface and \textsc{Hdt}.
For baselines, we uniformly set $p\text{=}10$k to boost their answer presence.
%(Sec.~\ref{subsec:kb-model}).
% Facts with external IDs, URLs, language tags and redundant schema information are removed.
To build
% lists based on
%lexical 
term
matching lists of question terms 
%among 
against
KB items, we used Elasticsearch~\cite{gormley2015elasticsearch}.
We use Wikipedia2Vec~\cite{yamada2020wikipedia2vec}
% , that incorporates signals from both textual co-occurrence and graph structure,
to compute embeddings for terms and KB items wherever needed.
Questions were segmented into phrases like \phrase{Harry Potter} and \phrase{theme music} using named entity recognition~\cite{ferragina2010tagme}.
%(alternatives could be query segmentation via corpus statistics~\cite{hagen2011query}).
% To treat popular multiword names like \phrase{Diego Maradona} as one question token, named entity recognition (NER) was applied on all questions~\cite{ferragina2010tagme}.
The depth of
% fetching from
the term-matching lists was set to $d\text{=}20$, and hyperparameters were tuned via dev sets to $h_{coh}\text{=}0.1$, $h_{conn}\text{=}0.3$,
% $h_{dens}=0.2$,
$h_{rel}\text{=}0.2$, $h_{match}\text{=}0.4$ for both benchmarks.
% discuss pruning threshold p here
% We use two values of the pruning threshold $p$, $10k$ and $100$.
The default setting for \clocq is an automatically chosen $k$
% (Sec.~\ref{subsec:dyn}; since $d\text{=}20$, $k \in [1, 5]$)
and $p\text{=}1$k
(Sec.~\ref{subsec:dyn}).
Since $d\text{=}20$, we have $k \in [1, 5]$.
This default configuration is implied when writing just ``\clocq''. % , if not stated otherwise.
% Since, $d = 20$ we have k $\in
% [1,\floor{log_2(20)}+1]=
%  here.
% We investigate effects of variation in $k$ and $p$ in our experiments. 
% We set $p = 10k$ as default.

% in our experiments.
% Analogously, since facts containing popular predicates (like \struct{spouse}) can often be extremely numerous,
% %currently
% % in \clocq
% we only consider predicates with $\leq p$ facts.
% % from the top-$k$ tuples.

\vspace*{-0.2cm}
\section{Results and insights}
\label{sec:exp-res}

\subsection{Key findings}
\label{subsec:key}

% We present
Our main results
on search space reduction
are in Table~\ref{tab:main-res}. 
%Note that the initial answer search space is comprised of entities and literals in the whole KB, and has a size of $\simeq 152M$. 
As a reference point, the
a-priori
% apriori 
answer search space consists of all entities and literals in the whole KB $K$, a total of about $152$M items.
% Statistical significance  top-$k$ w.r.t. the strongest baseline on each dataset.
% Key insights are presented below.
% \squishlist
    % \item \textbf{\clocq is more effective at search space reduction:}

\myparagraph{\clocq 
%is better at retaining 
keeps more
answers in its search space}
% In the default setting,
% We observed that answer presence for \clocq (default) is substantially better than all baselines. % for all pairwise comparisons.
% We observe that
\clocq outperforms the best baseline on answer presence for both benchmarks: by $5.8\%$ for LC-QuAD, and by $7.2\%$ for ConvQuestions, pushing the upper bound for performance of QA systems.
\clocq is able to keep $82.6\%$ (LC-QuAD) and $84.7\%$ (ConvQuestions) answers in its search space,
which is statistically significant for all pairwise comparisons with \textsc{Elq}
% ($76.7\%$ and $77.5\%$) 
and \textsc{Tagme},
% ($76.8\%$ and $69.1\%$) 
the strongest baselines for this task.
% The runtime is comparable to the $k\text{=}1$ setting, and the search space size much smaller.
% Already by computing
Importantly, $\clocq$ %does this with
% competitive
achieves this in
sub-second 
runtimes,
 %   completion times, 
% with the smallest runtime.
slightly faster than \textsc{Elq}, the fastest baseline. 
% $|\mathcal{S}|$ is slightly higher than for \textsc{Elq}, but smaller than for \textsc{Tagme}.
% with % top-1 results,
While \clocq (default) performs best, 
we note that \clocq ($k\text{=}1$)
% \clocq is able to keep $79.4\%$ (LC-QuAD) and $78.7\%$ (ConvQuestions) answers in its search space, outperforming \textsc{Elq} ($76.7\%$ and $77.5\%$) and \textsc{Tagme} ($76.8\%$ and $69.1\%$), the strongest baselines for this task.
% $\clocq$'s top-1 results are statistically significant for almost all pairwise comparisons
achieves an answer presence that is substantially better than that of all baselines as well,
showing the effectiveness of KB-aware signals for this task.
    % We explain the success of \clocq below.
    % : being faster than , on LC-QuAD, and slightly slower than ELQ on ConvQuestions.
    % However, this is closely connected to the size of the search space (details below): retrieving larger candidates translates to more time
% This shows the effectiveness of our mechanism for choosing $k$ dynamically.

\begin{figure*} [t]
    \begin{subfigure}[b]{0.3\textwidth}            
    \begin{center}
         \includegraphics[width=0.9\textwidth,height=0.17\textheight]{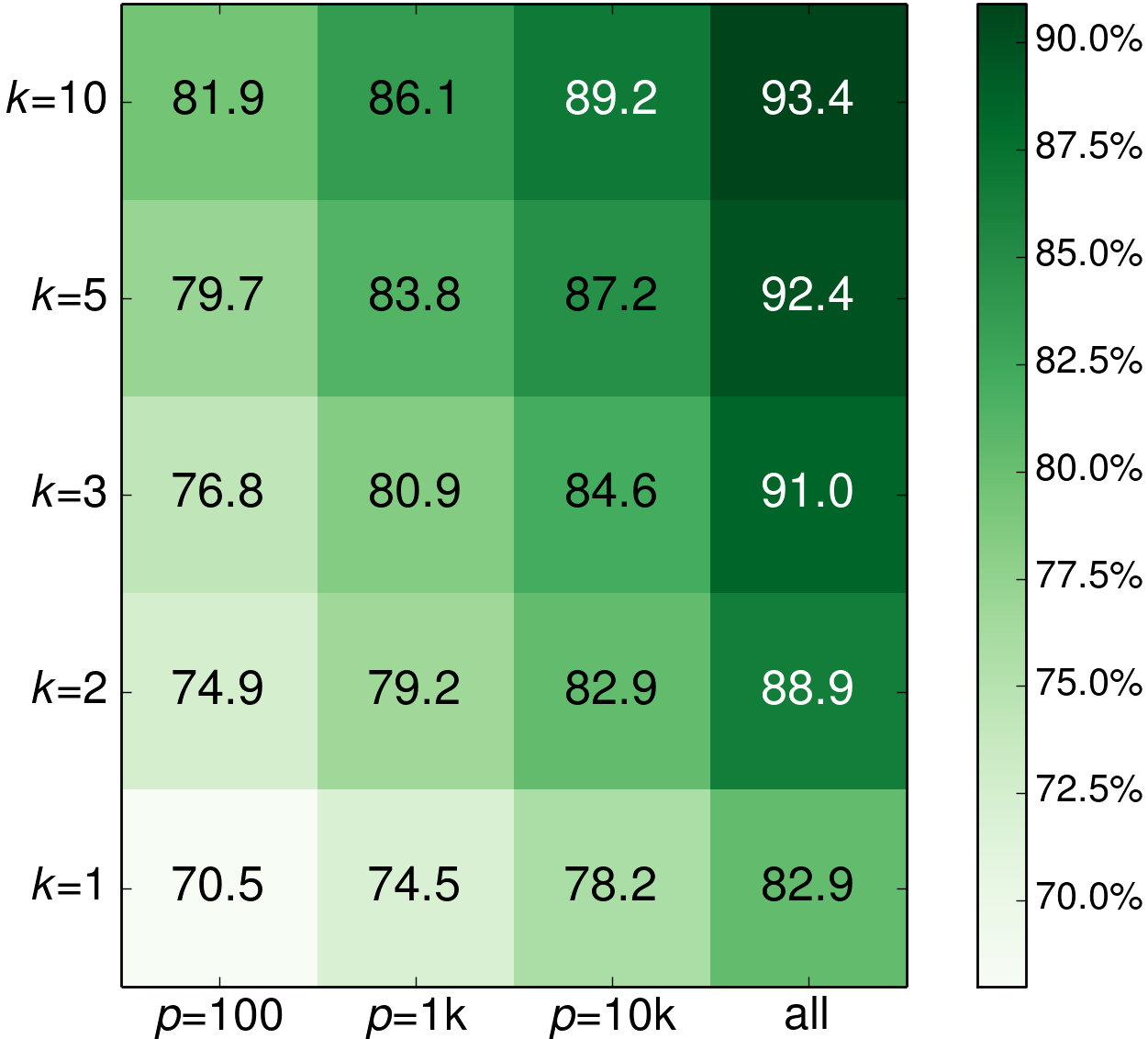}
         \caption{LC-QuAD: Answer presence.}
         \label{fig:heatmap-1}
         \end{center}           
    \end{subfigure}
    \begin{subfigure}[b]{0.3\textwidth}
         \begin{center}
         \includegraphics[width=0.9\textwidth,height=0.17\textheight]{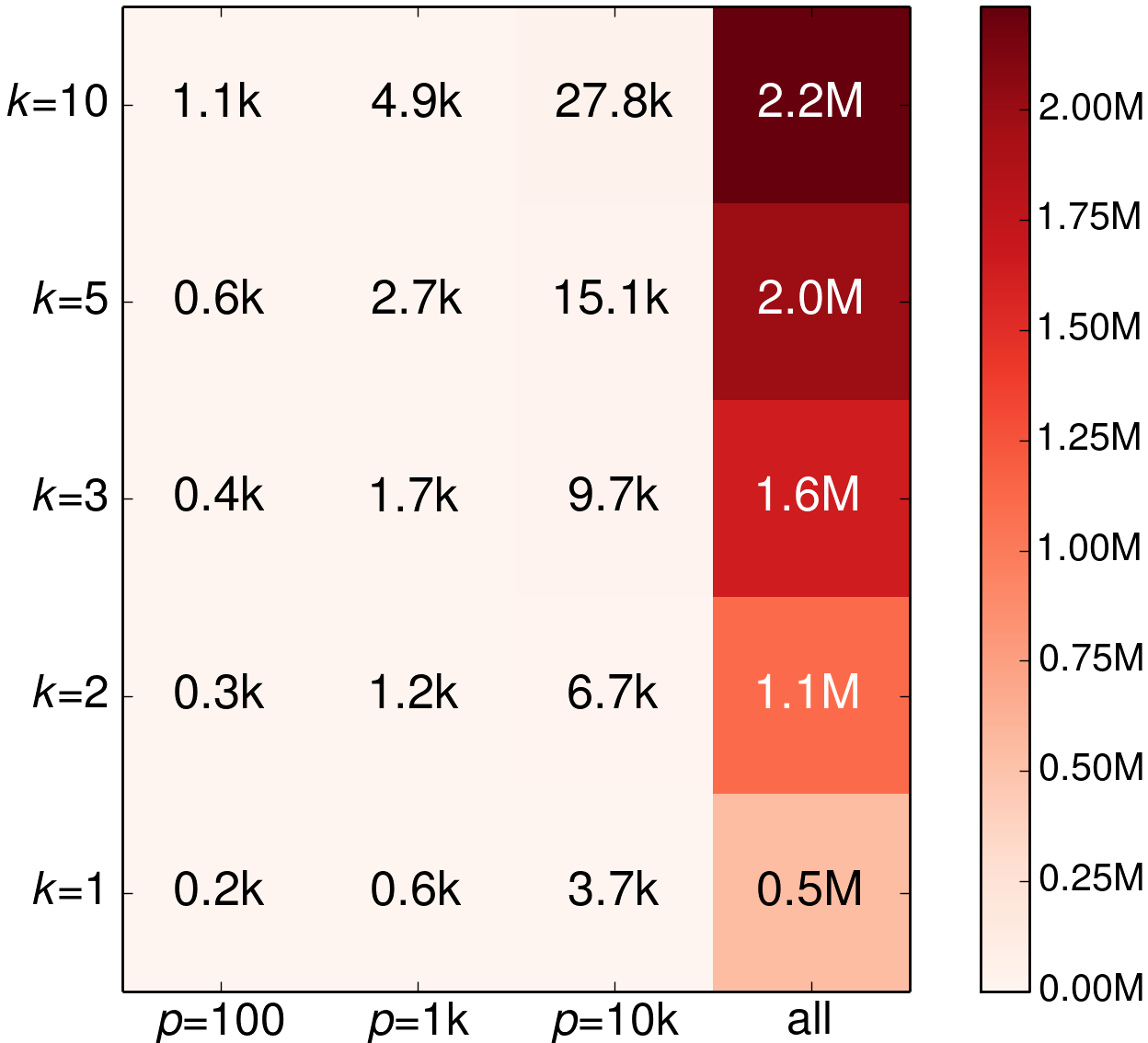}
        %  \caption{Search space size.}
         \caption{LC-QuAD: Search space size.}
         \label{fig:heatmap-2}
         \end{center}
    \end{subfigure}
    \begin{subfigure}[b]{0.3\textwidth}
         \begin{center}
         \includegraphics[width=0.9\textwidth,height=0.17\textheight]{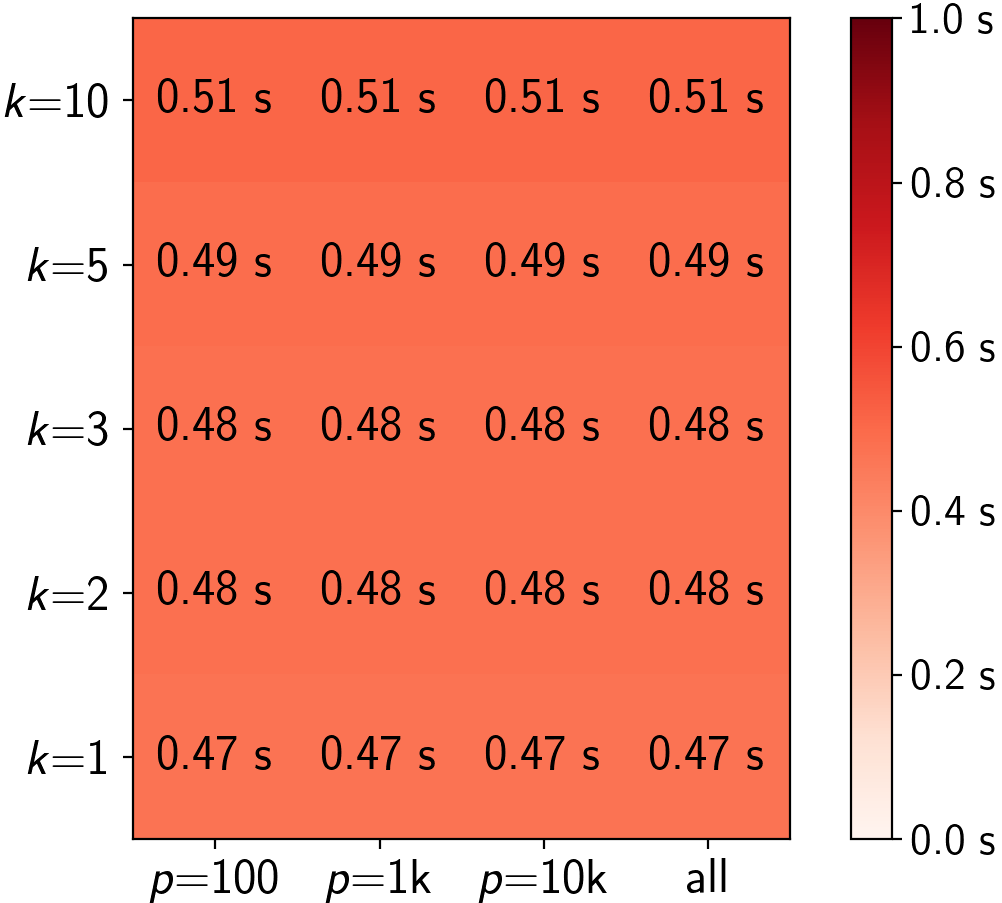}
         \caption{LC-QuAD: Runtimes.}
         \label{fig:heatmap-3}
         \end{center}
    \end{subfigure} %\\
    \begin{subfigure}[b]{0.3\textwidth}
         \begin{center}
         \includegraphics[width=0.9\textwidth,height=0.17\textheight]{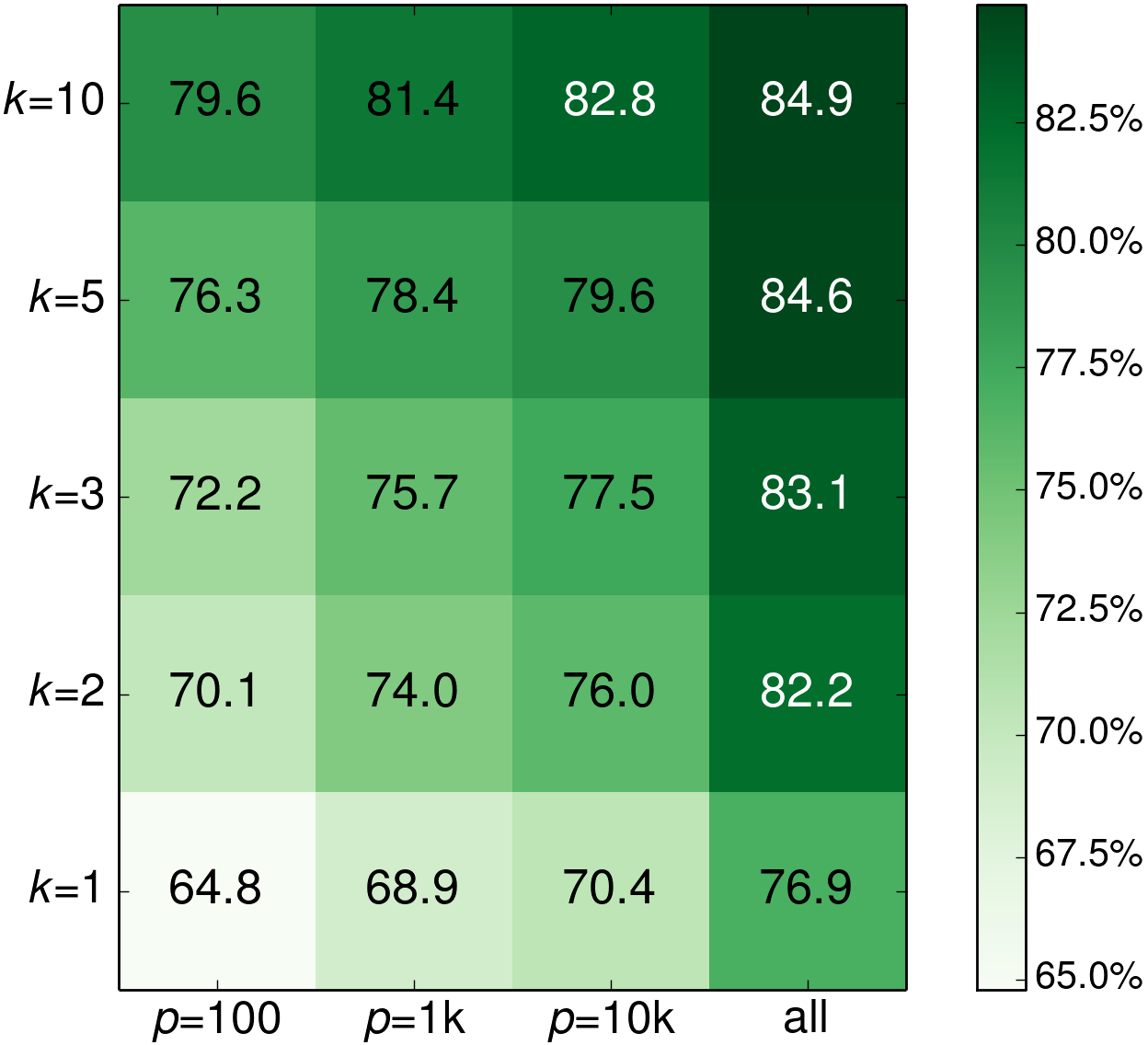}
         \caption{ConvQuestions: Answer presence.}
         \label{fig:heatmap-4}
         \end{center}
    \end{subfigure}
    \begin{subfigure}[b]{0.3\textwidth}
         \begin{center}
         \includegraphics[width=0.9\textwidth,height=0.17\textheight]{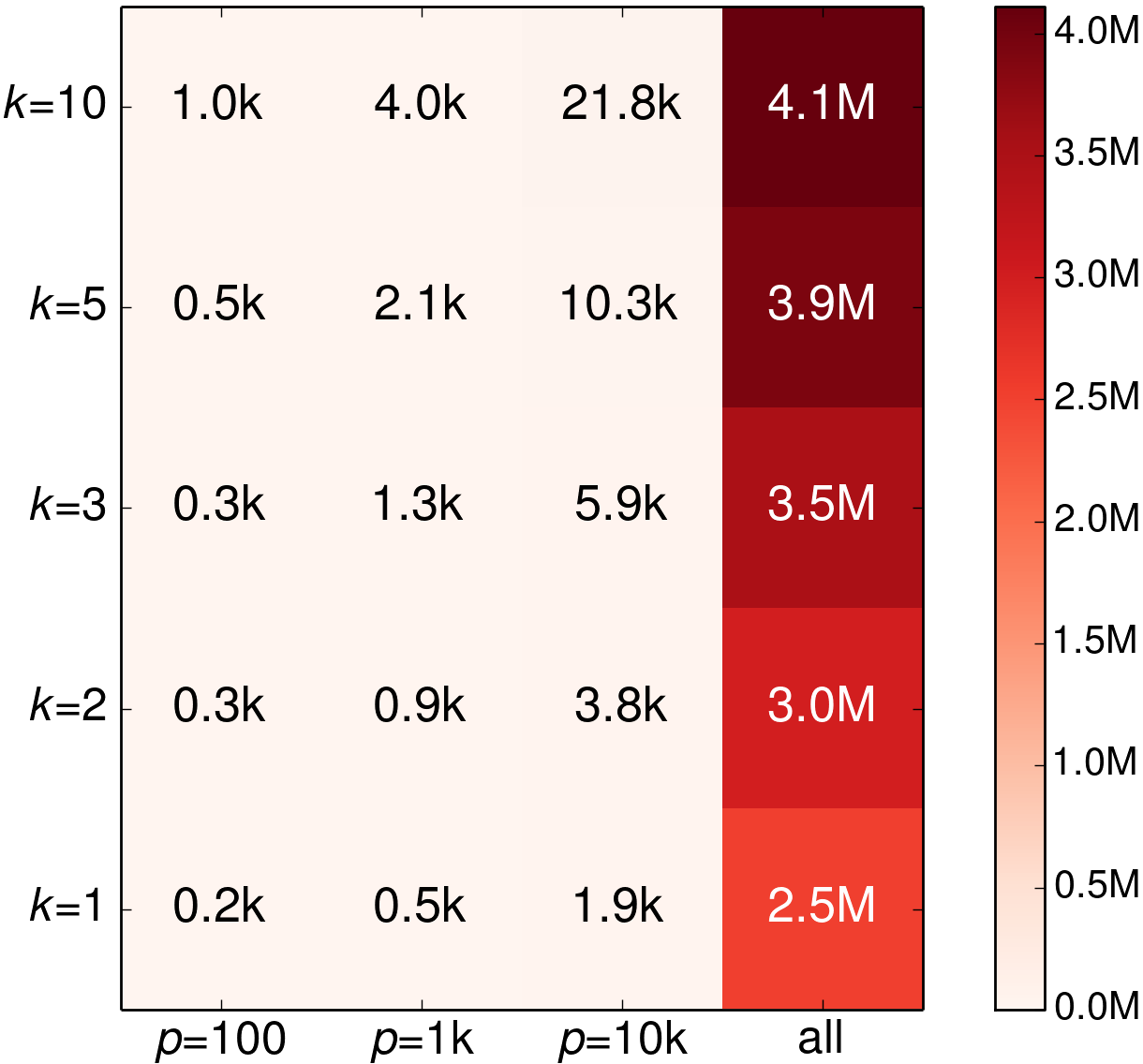}
         \caption{ConvQuestions: Search space size.}
         \label{fig:heatmap-5}
         \end{center}
    \end{subfigure}
    \begin{subfigure}[b]{0.3\textwidth}
         \begin{center}
         \includegraphics[width=0.9\textwidth,height=0.17\textheight]{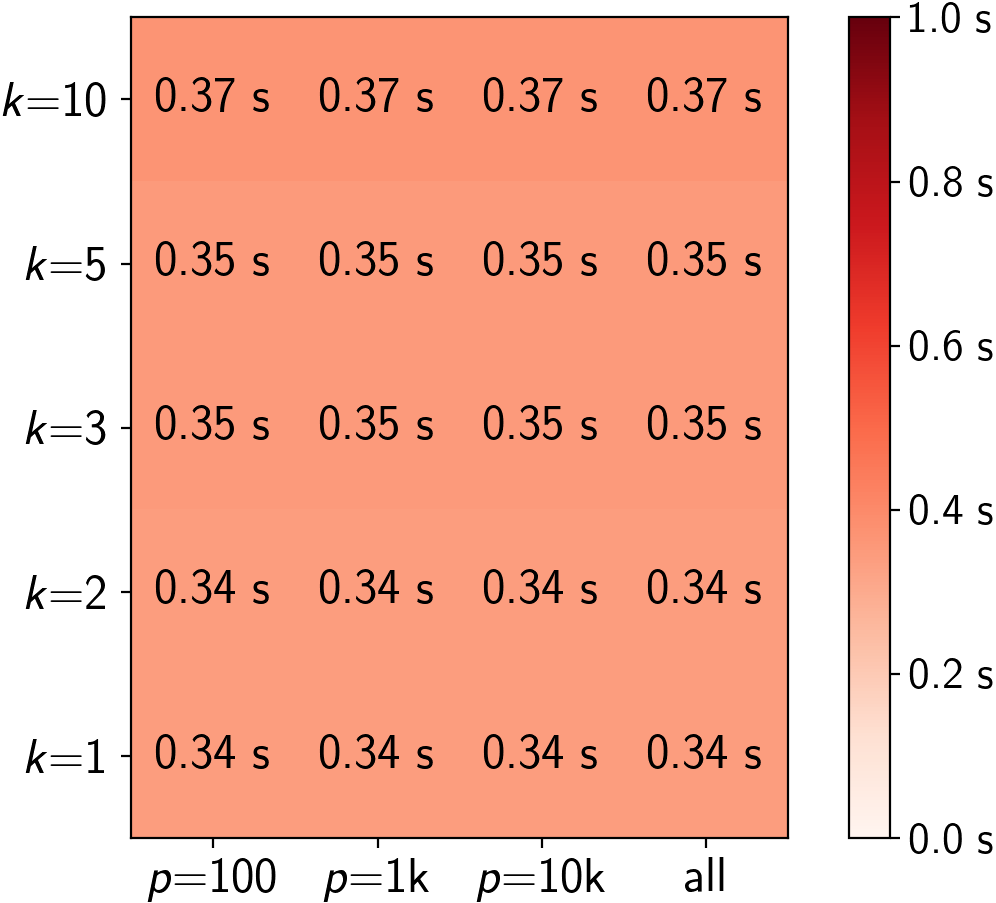}
         \caption{ConvQuestions: Runtimes.}
         \label{fig:heatmap-6}
         \end{center}
    \end{subfigure}\\
    \vspace*{-0.3cm}
    % \caption{Top-$k$ \clocq on the LC-QuAD dev set. Trends on the ConvQuestions dev set are the same.}
    \caption{Varying \clocq parameters on the LC-QuAD and ConvQuestions dev set.}
    \label{fig:heatmaps} 
    \vspace*{-0.5cm}
\end{figure*}

\myparagraph{Top-$\boldsymbol{k}$ 
results
%combinations 
add value over top-1}
%are better than top-1} 
% Since many NED methods 
%disambiguate only to the top-1 entity, 
% compute only top-1 entities,
% we use $k=1$ for \clocq for fairness in Table~\ref{tab:main-res}. However, 
%we believe that 
The true power of \clocq comes from the flexibility of top-$k$ outputs, coupled with the pruning threshold $p$.
Fig.~\ref{fig:heatmaps} shows variation in answer presence, search space size and runtime with $k$ and $p$ on the dev sets.
% The key point to note is
We see that
by increasing $k$ from $1$ to $10$,
\clocq achieves very good answer presence (going above $80\%$, Fig.~\ref{fig:heatmap-1} and Fig.~\ref{fig:heatmap-4}),
while keeping a tight threshold on
% facts
items
admitted into the search space (columns 1 and 2, $p$=100 or 1k).
%Note that, correspondingly, the context size 
% In this regime,
Here, the search space stays fairly small,
in the order of a few thousand KB items
%does not grow very large 
(Fig.~\ref{fig:heatmap-2} and Fig.~\ref{fig:heatmap-5}).
%just about $1k$ items.
% , instead of the $\simeq 4k$ for the % unrestricted top-1 scenario (where $p = 10k$ for \clocq and baselines).
    % reported in Table~\ref{tab:main-res} (column ``all'').
%    Thus, for deployment of \clocq, a practical
%    % choice
%    trade-off would be to have a reasonably high $k$
%    % ($\simeq 5$)
%    and suitably tight $p$. % ($\simeq 1000$)
%    % that is best for
%    % the answer presence-search space size
%    % this trade-off.
If, on the other hand, a QA system requires very high recall,
% % and hence very high answer presence,
% % rishi: equivalent to recall, hence commented
\clocq can achieve this
% % as well
by increasing $p$ (columns 3 and 4 in Fig.~\ref{fig:heatmap-1}/\ref{fig:heatmap-4}):
answer presence is well above $90\%$ and $80\%$, respectively. 
% Though at the cost of a much larger search space.
The price is a much larger search space.
Another observation is that due to the use of our efficient top-$k$ architecture and novel KB index, the timings are fairly stable when increasing $k$ and $p$. 
For change in $p$, we did not observe any increase in runtimes, and for $k$, the increase is $\leq 0.04$ seconds.
% These trends are almost the same for ConvQuestions.
% This will increase the 
% search space
% increases.
% to more than 1M items,
% but even this
% is
% % can still be
% tractable in terms of both memory consumption and runtime.
We added one top-$k$ variant with a good trade-off on the dev-set
($k\text{=}5$, $p\text{=}100$)
to Table~\ref{tab:main-res}. 
This significantly outperforms all baselines w.r.t. answer presence and runtime, with a very small search space size of only about $800$ items (last row). Among our baselines, \textsc{Earl}~\cite{dubey2018earl} can produce top-$k$ disambiguations: using $k\text{=}5$
% and $p=1$k
for \textsc{Earl} (fourth row) also increases its answer presence,
% from $53.8\%$ to $63.8\%$ (LC-QuAD)
but this is far below that of \clocq. % 's performance. % of \clocq.

We identify a \textit{trade-off} between answer presence and search space size as a major consideration for QA.
The best setting for $k$ and $p$ highly depends on the QA system operating on the contracted search space.
In general, for improving the answer presence, we recommend increasing $k$ rather than $p$.
Even though increasing both $k$ and $p$ cannot decrease the answer presence, the additional facts admitted into $\mathcal{S}$ could still distract the QA system and lead to longer runtimes.
Therefore, the choice of $k$ and $p$ depends on the maximum search space size and potential disambiguations per mention (manifested as $k$) a specific QA system can handle. %, without facing a significant runtime overhead or level of distraction. 

\begin{table} \small
    \caption{Anecdotal examples
    from test sets of considered benchmarks
    for which only \clocq had an answer in the search space (green phrases denote correct mappings).} % The first three are from LC-QuAD, and last three from ConvQuestions.}
    \vspace*{-0.3cm}
    % \resizebox*{\columnwidth}{!}{
    	\begin{tabular}{p{8cm}}
    		\toprule
    		    \textbf{Robust w.r.t.
        		wrong mappings and redundant question words}\\
        		\midrule
        % 		\ctext[RGB]{0,209,0}{\utterance{How is the main soccer club of the german city D\"usseldorf called?} (ConvQuestions)}\\
                \anecdote{\utterance{How is the main soccer club of the german city D\"usseldorf called?} (ConvQuestions)}\\
        		\underline{\clocq}: \\
        		\phrase{main}$\mapsto$\itemslist{Frankfurt (Main), Main (river), Offenbach am Main};\\
        		\phrase{soccer}$\mapsto$\itemslist{football, \textcolor{OliveGreen}{Football team}};\\
        		\phrase{club}$\mapsto$\itemslist{Nightclub, Torino F.C.};\\
        		\phrase{german}$\mapsto$\itemslist{\textcolor{OliveGreen}{German}, German Empire};\\
        	    \phrase{city}$\mapsto$\itemslist{Manchester City F.C., Birmingham City F.C., Stoke City F.C., Cardiff City F.C.};\\
        	    \phrase{D\"usseldorf}$\mapsto$\itemslist{\textcolor{OliveGreen}{D\"usseldorf, Fortuna D\"usseldorf}};\\
        		\underline{\textsc{Tagme}}: \\
        		\phrase{main}$\mapsto$\itemslist{Main (river)};\\
        		\phrase{soccer}$\mapsto$\itemslist{football};\\
        		\phrase{club}$\mapsto$\itemslist{sports club};\\
        		\phrase{german}$\mapsto$\itemslist{Germany};\\
        		\phrase{city}$\mapsto$\itemslist{City of London};\\
        		\underline{\textsc{Elq}}: \\
        % 		\phrase{main}$\mapsto$\itemslist{Klaus~Barbie};\\
        		\phrase{german}$\mapsto$\itemslist{Germany};\\
        		\phrase{soccer}$\mapsto$\itemslist{football};\\
    		    
        		% \textbf{Question}	\\ \toprule
        		%% LCQUAD:
        % 		\utterance{Who used tuberculosis as a murder weapon and was a sculptor of a pressure volume diagram?}\\
                \midrule
                \textbf{Automatic top-$k$ for all question words can cover for errors}\\
                \midrule
                \anecdote{\utterance{Who is the son of the brother of Queenie Padilla?} (LC-QuAD)}\\
                \underline{\clocq}: \\
        		\phrase{son}$\mapsto$\itemslist{Son en Breugel, \textcolor{OliveGreen}{nephew}, Mae Hong Son, Porto do Son};\\
        		\phrase{brother}$\mapsto$\itemslist{sibling};\\
        		\phrase{Queenie Padilla}$\mapsto$\itemslist{\textcolor{OliveGreen}{Queenie Padilla}};\\
        		\underline{\textsc{Tagme}}: \\
        		\phrase{Who}$\mapsto$\itemslist{World Health Organization};\\
        		\phrase{brother}$\mapsto$\itemslist{Brother};\\
        		\underline{\textsc{Elq}}: \\ 
        		\phrase{Padilla}$\mapsto$\itemslist{Zsa Zsa Padilla}
        		\\ \midrule
        		
        % 		\textbf{Top-$k$ for all question words can cover for errors}
        % 		\\ \midrule
        % 		\anecdote{\utterance{Who was the director of the movie The Avengers?} (ConvQuestions)}\\
        % 		\underline{\clocq}: \\
        % 		\phrase{director}$\mapsto$\itemslist{\textcolor{OliveGreen}{film director}, cinematographer};\\
        % 		\phrase{movie}$\mapsto$\itemslist{\textcolor{OliveGreen}{film}, film producer};\\
        % 		\phrase{The Avengers}$\mapsto$\itemslist{\textcolor{OliveGreen}{The Avengers(2012 film)}, The Avengers(1998 film), The Avengers(1950 film), Avengers(comic)};\\
        % 		\underline{\textsc{Tagme}}: \\
        % 		\phrase{Who}$\mapsto$\itemslist{The Who};\\
        % 		\phrase{director}$\mapsto$\itemslist{\textcolor{OliveGreen}{film director}};\\
        % 		\phrase{The Avengers}$\mapsto$\itemslist{\textcolor{OliveGreen}{film}, The Avengers(1998 film)};\\
        % 		\underline{\textsc{Elq}}: \\
        % 		\phrase{The Avengers}$\mapsto$\itemslist{The Avengers(1998 film)};
        % 		\\ \midrule
        		
        		\textbf{Auto-$k$ mechanism can identify highly ambiguous words}
        		\\ \midrule
        		\anecdote{\utterance{Who is the composer of All We Know?} (LC-QuAD)}\\
        		\underline{\clocq}: \\
        		\phrase{composer}$\mapsto\langle$\struct{\textcolor{OliveGreen}{composer}, film score composer}$\rangle$; \\
        		\phrase{All We Know}$\mapsto\langle$\struct{\textcolor{OliveGreen}{All We Know (Paramore)}, For All We Know (album), All We Know (Chainsmokers), For All We Know (Carpenters), For All We Know (1934 song)}$\rangle$; \\
        		\underline{\textsc{Tagme}}: \\
        		\phrase{Who}$\mapsto$\itemslist{The Who};\\ \phrase{composer}$\mapsto$\itemslist{\textcolor{OliveGreen}{composer}}; \\
        		\phrase{All We Know}$\mapsto$\itemslist{For All We Know (Carpenters)};\\
        		\underline{\textsc{Elq}}: \\
        		\phrase{All We Know}$\mapsto$\itemslist{All We Know (Chainsmokers)};
        		\\ \midrule
        		
        % 		\utterance{What type of film is significant for William McDonough?}\\
        %         \utterance{what was the name of Dorethea Lange's spouse in the year 1920?} (LC-QuAD)\\ \midrule
        %         \utterance{Name the Sparkasse with highest tier 1 capital ratio?}\\
        %         \utterance{Which is the railway tunnel with terminus Erstfeld and contains the word base in it's name?} \\ \midrule
    	        %% CONVQUESTIONS
        % 		\utterance{What does the spouse of Thomas Müller, Lisa Trede-Müller, work as in 2018?} (ConvQuestions) \\
        % 		\utterance{How many children did Morticia Addams have in The Addams Family movie?}\\
        % 		\utterance{How is the main soccer club of the german city Düsseldorf called?}\\
        % 		\utterance{What is the name of the first book in Bill Hodges trilogy?}\\
        % 		\utterance{What are the names of the two lead male characters in the book Of mice and men?}\\
        % 		\utterance{In Star Trek: The Next Generation, the ship is named what?}\\
                \textbf{KB connectivity is a vital indicator for understanding context}
        		\\ \midrule
                \anecdote{\utterance{Who was the screenwriter for Crazy Rich Asians?} (ConvQuestions)}\\
                \underline{\clocq}: \\
        		\phrase{screenwriter}$\mapsto$\itemslist{\textcolor{OliveGreen}{screenwriter}};\\
        		\phrase{Crazy Rich Asians}$\mapsto$\itemslist{\textcolor{OliveGreen}{Crazy Rich Asians (film)}};\\
        		\underline{\textsc{Tagme}}:\\
        		\phrase{Crazy Rich Asians}$\mapsto$\itemslist{Crazy Rich Asians (book)};\\
        		\underline{\textsc{Elq}}: \\
        		\phrase{Crazy Rich Asians}$\mapsto$\itemslist{Crazy Rich Asians (book)};\\
        % 		\utterance{What actor played the role of Jason Bateman’s older brother in the sitcom Arrested Development?} (ConvQuestions)\\
        % 		\utterance{What sporting venue hosted the first Major League Soccer championship game in 1996?}\\
    		\bottomrule
    	\end{tabular} %}
% 	First three are from ConvQuestions-FQ, last three from LC-QuAD2.0-CQ.
	\label{tab:anec}
% 	\vspace*{-0.3cm}
%%%GW: don't place the tables too tight to each other;
%%% if there is space problem, fix this elsewhere (i.e., trim wording: more concise is better anyway)
\end{table}

% EARL Top-k on ConvQuestions:  53.4 % answer presence, 2023 items
% EARL Top-k on LCQuAD2.0:  65.9 % answer presence, 2192 items
\noindent

% \myparagraph{\clocq for QA} \clocq outputs
% % (k $= 5$, p $= 100$)
% for LC-QuAD fed into the popular KB-QA system GRAFT-Net~\cite{sun2018open} led to significantly better answering performance (P@1 = $0.154$, $0.71$ second faster)
% % (MRR = $0.214$)
% than when \textsc{Tagme} (best baseline on LC-QuAD) (P@1 = $0.138$)
% %, MRR = $0.231$)
% was used instead.
% % Systematic investigations of integrating \clocq with downstream QA models is left as future work.

\myparagraph{Impact on KB-QA} 
While answer presence is an important measure
% giving
creating an upper bound for the QA system,
the key goal of this work is to enhance the performance on the downstream QA task.
To study these effects, we feed the outputs of \clocq and the baselines
% (facts and KB items)
into the popular KB-QA system GRAFT-Net~\cite{sun2018open} and ran the two benchmark suites. 
% For \clocq we use the dynamic k variant with $p=1k$, to avoid fine-tuning of \clocq-parameters specifically for usage in GRAFT-Net.
We report the standard QA metrics~\cite{saharoy2021question} \textit{Precision~at~1} (P@1), \textit{Mean reciprocal rank} (MRR) and \textit{Hit~at~5} (Hit@5).
Results are
% shown
in Table~\ref{tab:graftnet}. For LC-QuAD, the configuration with \clocq significantly outperforms the two strongest baselines on all metrics.
For ConvQuestions, \clocq has the best performance 
%on all metrics except P@1, for which \textsc{Elq} performs slightly better.
on MRR and Hit@5, and is only slightly behind \textsc{Elq} on P@1.
These results show the benefits of \clocq for downstream QA. 
% Also, 
% \GW{show all three run-times explicitly!!!}\\
\clocq generates the search space faster:
the average runtimes per query are
$0.49$ s
% (0.5 * 2000 + 0.42 * 300) / 2300 = 0.49 s (weighted avgs of runtimes on both benchmarks)
for \clocq,
% (0.62 * 2000 + 0.47 * 300) / 2300 = 0.60 s
$0.60$ s for \textsc{Elq}+\textsc{Hdt},
and
% (1.14 * 2000 + 1.43 * 300) / 2300 = 1.18 s
$1.18$ s for \textsc{Tagme}+\textsc{Hdt}.
%$0.11$ s and $0.71$ s for \textsc{Elq}+\textsc{Hdt} and \textsc{Tagme}+\textsc{Hdt}, respectively) 
%irectly affects the responsiveness of the QA system.
%%%GW: the following point is made in Discussion later on, hence commented out here
%owever, we notice a gap between the upper bounds of the search spaces and the performance of the QA system, that calls for further research to make full use of the search space.

% \clocq outputs for LC-QuAD fed into the popular KB-QA system GRAFT-Net~\cite{sun2018open} led to significantly better answering performance (P@1 = $0.154$, $0.71$ second faster) than when \textsc{Tagme} (best baseline on LC-QuAD) (P@1 = $0.138$) was used instead.

\begin{table} \small
    \caption{Impact of \clocq on KB-QA.}
    \vspace*{-0.3cm}
    \newcolumntype{G}{>{\columncolor [gray] {0.90}}c}
    % \resizebox*{\columnwidth}{!}{
    	\begin{tabular}{l G G G c c c}
    		\toprule
    		    \textbf{Benchmark}	 & \multicolumn{3}{G}{\textbf{LC-QuAD2.0-CQ}}   & \multicolumn{3}{c}{\textbf{ConvQuestions-FQ}}  \\
    		    \midrule
    		    \textbf{QA system $\rightarrow$}	 & \multicolumn{3}{G}{\textbf{GRAFT-Net~\cite{sun2018open}}}  & \multicolumn{3}{c}{\textbf{GRAFT-Net~\cite{sun2018open}}} \\ 
        		\textbf{Search space $\downarrow$}	 & \textbf{P@1}  & \textbf{MRR} & \textbf{Hit@5} & \textbf{P@1}  & \textbf{MRR} & \textbf{Hit@5} \\ 
        		\toprule
        		\textbf{\clocq}		                    &	$\boldsymbol{0.197}$*	&	$\boldsymbol{0.268}$* &	$\boldsymbol{0.350}$* &	$0.207$	&	$\boldsymbol{0.264}$   &	$\boldsymbol{0.337}$   \\
        		\textbf{\textsc{Elq}+\textsc{Hdt}}		&	$0.168$	&	$0.224$ &	$0.288$ &	$\boldsymbol{0.213}$   &	$0.256$   &	$0.313$   \\
        	    \textbf{\textsc{Tagme}+\textsc{Hdt}}	&	$0.171$	&	$0.225$ &	$0.291$ &	$0.167$	&	$0.204$   &	$0.237$   \\
    		\bottomrule
    	\end{tabular} %}
    \label{tab:graftnet}
    \vspace*{-0.3cm}
\end{table}

% \vspace*{-0.3cm}
\subsection{In-depth analysis}
\label{subsec:analysis}

\myparagraph{\clocq 
%disambiguates more 
identifies relevant concepts and types}
    % \clocq has larger search spaces than most baselines:
%On closer inspection of several anecdotal cases,
%we found that 
    For many questions, \clocq identifies not just additional entities but also concepts and types that are missed by baselines.
    Since $k$>1 trivially adds more
    % (related)
    KB items,
    we set $k\text{=}1$ for fair comparison
    in this analysis.
    % in these experiments.
    % Including facts with these items amounts to the larger search space that we see for $\clocq$ in Table~\ref{tab:main-res}.
    For example, in \utterance{What was the name of the theme music for the television series Mash?}, \textsc{Elq} disambiguates only \phrase{Mash} (incorrectly), to the 1970 film.
    % with the same name.
    \clocq, on the other hand, finds:
    % the following:
    \phrase{name} $\mapsto$ \struct{personal name}, \phrase{theme music} $\mapsto$ \struct{theme music}, \phrase{television series} $\mapsto$ \struct{television series}, and \phrase{Mash} $\mapsto$ \struct{M*A*S*H} (the TV series, correct).
    % using($k$>1 would trivially add more (relevant) items)
    On average, \clocq finds $4.68$ KB items per question (LC-QuAD), while \textsc{Elq}, \textsc{Aida}, and \textsc{Tagme} find $1.82$, $2.65$ and $3.75$, respectively.
    We verified
    % in experiments
    that these additionally disambiguated types and concepts help:
    % in retaining the answer:
    when
    % such types and concepts
    % these are 
    removed from 
    \clocq's 
    % the
    output, answer presence drops from $78.3\%$ to $65.5\%$ (LC-QuAD dev).
    %, on the LC-QuAD2.0-CQ dev set.
    Note that
    % standard
    standalone
    NED evaluation % (with micro-averaged precision and recall)
    is out of scope here, because QA benchmarks have no ground-truth for KB item disambiguation.
    %, being limited to gold answers for questions.
    %there is no full ground truth annotations in QA benchmarks.
    %However, this is also not the primary evaluation focus here. %in this work.
    %
    % \item \textbf{Item-wise \clocq outperforms naive variant:} Significantly better answer presence of item-wise TA illustrates the value of incorporating additional signals into the tuple ranking process. Nevertheless, competitive values for the naive version ($76.2\%$ and $72.9\%$) operating with only lexical matches, w.r.t. sophisticated neural baselines show the power of simple yet elegant top-$k$ methods for this task.

%    lower than the $82.7\%$ for \clocq in the same setting (search spaces
    % sizes
%    comparable at $\simeq 2.8k$).

\myparagraph{Representative examples}
Representative examples
of success cases for \clocq
are in Table~\ref{tab:anec}.
In the first example, the inherent focus of \clocq on related concepts leads to some incorrect disambiguations:
given the football context, \phrase{city} is disambiguated to a set of football clubs.
Further, \phrase{main} appears to be the German river.
Despite this noise \clocq, is able to correctly detect \struct{D\"usseldorf} within its disambiguations. 
Interestingly, the correct answer \struct{Fortuna D\"usseldorf} is also found, taking the football context into account.
In the second example, \phrase{son of the brother} is not correctly disambiguated in top-$1$ results, but leveraging the Auto-$k$ mechanism, the method can make up for this error and add the correct KB item (\struct{nephew}).
Similarly, for the next question \clocq identifies \phrase{All We Know} as a highly ambiguous phrase, and returns top-$5$ disambiguations. 
The baselines \textsc{Tagme} and \textsc{Elq} fail to get the correct entity on the top rank, which means that the question is a lost cause.
In the last example, KB connectivity helps \clocq:
% to resolve the ambiguity:
both baselines identify a book, while the question word \phrase{screenwriter} gives a clear hint that the
question is about a movie.
\clocq disambiguates \struct{screenwriter} and leverages KB connectivity
to disambiguate the correct movie (\struct{screenwriter} and \struct{Crazy Rich Asians (film)} are connected in 1-hop).
% none of the songs disambiguated by \textsc{Tagme} or \textsc{Elq} has a composer.
% \clocq leverages the information that the intended song has a connection with the \struct{composer} predicate in the KB,
% and detects the correct song.
% Similarly, \phrase{Crazy Rich Asians} could refer to the book or movie.
% The screenwriter predicate gives a clear hint, which helps \clocq in resolving the ambiguity.

\myparagraph{Ablation studies}
\clocq includes four signals in its architecture: this naturally calls for
% a systematic ablation study.
ablations (Table~\ref{tab:abl-study}, dev sets).
% The result of our ablation study (for our default setting, dynamic $k$ and $p\text{=}10$k) is shown in Table~\ref{tab:abl-study}.
% We found that
% performance
% rishi: performance is ambiguous
Answer presence
on ConvQuestions
% dev set
dropped for each single signal that is removed,
% respectively,
showing that all four
% signals
matter (* = significant drop from full configuration).
On LC-QuAD,
% dev
trends are similar, just that removing relevance led to a slightly increased answer presence.
% details already in table caption
% , with McNemar's test, $\mathrm{p} < 0.05$).).
% Results are similar for the ConvQuestions benchmark.
% For ConvQuestions,
% % dev set
% performance dropped from $71.7\%$ (full) to $66.0\%$* (no $coh$ + $conn$), $46.4\%$* (no $rel$ + $match$), $68.9\%$* (no $coh$), $68.6\%$* (no $conn$), $70.7\%$ (no $rel$), and $60.9\%$* (no $match$). 
%{\color{blue}
While removing a single component has only small influence,
dropping the pair of local and global signals (like $match + rel$, or $coh + conn$) often results in noticeable loss.
However, such choices may indeed need to be made when runtime is of utmost importance, since computing $coh$ and $conn$ are the most time-consuming steps in
% the
\clocq.

\begin{table} 
    \caption{Ablation study of configurations in \clocq.}
    % \vspace*{-0.3cm}
    \setlength{\tabcolsep}{4pt}
    \newcolumntype{G}{>{\columncolor [gray] {0.90}}c}
    \resizebox*{\columnwidth}{!}{
    	\begin{tabular}{l G G G c c c}
    		\toprule
    		\textbf{Benchmark}	 & \multicolumn{3}{G}{\textbf{LC-QuAD2.0-CQ}}   & \multicolumn{3}{c}{\textbf{ConvQuestions-FQ}}  \\
        		\midrule
    % 		\textbf{Metric $\rightarrow$}    & \textbf{Answer}   & \textbf{Context} & \textbf{Runtime}  & \textbf{Answer}   & \textbf{Context} & \textbf{Runtime} \\
    %     	\textbf{Method $\downarrow$}	        & \textbf{presence} & \textbf{size} &    & \textbf{presence} & \textbf{size}  & \\
    		 \textbf{Method $\downarrow$}	 & \textbf{Ans. pres.}  & \textbf{$|\mathcal{S}|$} &\textbf{Time}  & \textbf{Ans. pres.}  & \textbf{$|\mathcal{S}|$}  &\textbf{Time} \\
    		\toprule
	        \textbf{\clocq}                     &	$0.803$	&	$1.5$k &	$0.47$ s	&	$0.760$	&	$1.1$k &	$0.34$ s  \\ \midrule
    		\textbf{w/o $match$}		        &	$0.726$*	&	$1.3$k &	$0.46$ s	&	$0.607$*	&	$0.9$k &	$0.30$ s  \\
    		
    		\textbf{w/o $rel$}	            &  $0.806$	&	$1.5$k &	$0.47$ s	&	$0.746$*	&	$1.2$k &    $0.32$ s	\\
    		\textbf{w/o $conn$}               &	$0.790$*	&	$1.5$k &	$0.41$ s	&	$0.746$*	&	$1.2$k &	$0.26$ s	\\	
    		\textbf{w/o $coh$}		        &	$0.802$	&	$1.5$k &	$0.40$ s	&	$0.750$	&	$1.1$k &	$0.24$ s  \\
    		\textbf{w/o $match$ + $rel$}		&	$0.733$*	&	$1.3$k &	$0.47$ s 	&	$0.618$*	&	$0.9$k &	$0.30$ s  \\
    		\textbf{w/o $coh$ + $conn$}		&	$0.791$*	&	$1.5$k &	$0.34$ s	&	$0.743$*	&	$1.2$k &	$0.22$ s  \\
    		\bottomrule
    	\end{tabular}
    }
    \label{tab:abl-study}
    \vspace*{-0.3cm}
\end{table}

% respectively.
% A * indicates a significant drop from the full configuration, for the respective configuration (with McNemar's test, $\mathrm{p} < 0.05$).
% \GW{why is this on dev, why not test? add a concluding sentence (e.g., dropping coh+conn has a notable but far from dramatic loss -- so only using match+rel could be viable)?}

\myparagraph{Error analysis} \clocq
% (item-wise top-1)
% does not have
misses
the answer in $\mathcal{S}$
% its search space
just about $20\%$ of the time
(both benchmarks),
% . There are two %main sources of
arising from two error cases:
% kinds of errors:
i)
% for about half of the error cases,
the answer is missing
in the computed set of facts, as the depth-$d$
term matching does not retrieve the relevant items
(LC-QuAD $44.8\%$, ConvQuestions $46.7\%$);
% this could be mitigated by increasing $d$.
% , but would incur additional cost.
%This calls for more informed retrieval.
%when the answer is not contained in the returned set of facts:
% (any of the KB items that has the answer in its facts is not in the top-1):
%this happens in $44.8\%$ (LC-QuAD) and $46.7\%$
%%%GW: this is 44.8% of the error cases, could be misinterpreted
%(ConvQuestions) cases, suggesting that lexical matching falls short and more informed retrieval could help; 
% keep it simple
%and ii) the answer is there but the KB-item that has the answer in its set of facts does not appear in the top-1
% tuple:
%combination (rest of the cases).
% For the other half,
and ii) the answer is in the candidate space,
but the top-$k$ algorithm fails to return one or more relevant items
(LC-QuAD $55.2\%$, ConvQuestions $53.3\%$).
% this is for the rest of the cases (
% $55.2\%$ (LC-QuAD) and $53.3\%$ (ConvQuestions).
%This is a limitation of the top-$k$ architecture, and could be addressed by considering more signals, or a better aggregation function. %than a linear combination. 
Both cases could be countered by increasing $d$ or the range of $k$,
% This could be countered by increasing $k$
at the cost of increased end-to-end runtimes.

% \myparagraph{Manual error analysis.}

\myparagraph{Automatic choices for 
% $k$ and 
$\boldsymbol{p}$}
% for this investigation. % for an automatic setting of $k$ and $p$, on the dev sets.
% We show results for our default setting, with a higher $p$ ($p$=10k), and three variants with dynamic $k$ and $p$. 
Table~\ref{tab:dyn} shows results of various choices.
As discussed in Sec.~\ref{subsec:dyn}, $p$ can be set as $f(k)$. 
We tried $p\text{=}10^{5-k}$ first, and found that $p$ is falls off too drastically.
Therefore, we compared with smoother versions $p\text{=}10^{5-0.5 k}$ and $p\text{=}10^{4-0.5 k}$.
% , where the factor $0.5$ avoids the drastic drop of $p$ with increasing $k$.
Again, there is a trade-off between answer presence and search space size:
having $p\text{=}10^{5-0.5 k}$ gives the best answer presence, but $p\text{=}10^{4-0.5 k}$ has a much smaller
$|\mathcal{S}|$.
% search space size.
The runtime was almost the same across all variants. %, and therefore dropped for better readability.
Overall, we found a static setting of $p$ to perform slightly better with respect to the trade-off.
% For reference, we include results for
% fine-tuned
% tuned parameter settings
% (w.r.t. trade-off)
% for $k$ and $p$.
% Performance of
% A dynamic $k$ performs comparably on LC-QuAD and better on ConvQuestions, indicating the effectiveness of our % automatic
% proposed
% entropy-based mechanism for choosing $k$.
% for disambiguation.

\begin{table} \small
    \caption{Effect of choosing $k$ and $p$ dynamically per term.}
    % \vspace*{-0.3cm}
    \newcolumntype{G}{>{\columncolor [gray] {0.90}}c}
    \resizebox*{\columnwidth}{!}{
    	\begin{tabular}{l G G c c}
    		\toprule
    		    \textbf{Benchmark}	 & \multicolumn{2}{G}{\textbf{LC-QuAD2.0-CQ}}   & \multicolumn{2}{c}{\textbf{ConvQuestions-FQ}}  \\
        		\midrule
        % 		\textbf{\clocq}	 & \textbf{Answer}  & \textbf{Context}  & \textbf{Answer}  & \textbf{Context}  \\
        % 			    & \textbf{presence} & \textbf{size}     & \textbf{presence} & \textbf{size}\\
                \textbf{\clocq $\downarrow$}	 & \textbf{Ans. pres.}  & \textbf{$|\mathcal{S}|$}  & \textbf{Ans. pres.}  & \textbf{$|\mathcal{S}|$}  \\ \toprule
%         			    & \textbf{presence} & \textbf{size}     & \textbf{presence} & \textbf{size}\\        
%        		\toprule
        		{$k$=Auto, $p\text{=}1$k} (default)		            &  	$0.803$	&	$1.5$k &	$0.760$ &	$1.0$k \\ 
        		{$k$=Auto, $p\text{=}10$k}		        &  	$0.837$	&	$7.8$k &	$0.778$ &	$4.6$k \\ \midrule
        		{$k$=Auto, $p\text{=}10^{5-k}$}		    &  	$0.776$	&	$1.5$k &	$0.728$ &	$0.9$k \\
        		{$k$=Auto, $p\text{=}10^{5-0.5\dot k}$}	&  	$0.835$	&	$6.3$k &	$0.778$ &	$4.3$k \\
        		{$k$=Auto, $p\text{=}10^{4-0.5\dot k}$}   &  	$0.798$	&	$1.0$k &	$0.734$ &	$0.8$k \\
        		\midrule
                {$k$=3, p=1k}       &  	$0.809$	&	$1.7$k &	$0.757$ &	$1.3$k \\	
                {$k$=5, p=100}   &  	$0.797$	&	$0.6$k &	$0.763$ &	$0.5$k \\	
    		\bottomrule
    	\end{tabular} }
    \label{tab:dyn}
    \vspace*{-0.3cm}
\end{table}

\myparagraph{IR-based extension}
% Unlike for standard NED, in QA it is not of importance which specific KB items are disambiguated. 
% The focus is on the facts that are given as input to the downstream KB-QA system.
% Therefore,
An intuitive extension or alternative 
% following recent developments in QA over heterogeneous sources
is to fetch a larger subset of the KB, verbalize these facts~\cite{oguz2021unified,agarwal2020knowledge,pramanik2021uniqorn}, and use a standard IR pipeline to retrieve the most relevant facts for use by the QA system.
We implemented such a variant, treating the question as a query and the verbalized facts as the set of documents. BM25~\cite{robertson2009probabilistic} is used for scoring fact-relevance,
and returns the top-100 or top-1000 facts.
We used the {rank\_bm25} module\footnote{\url{https://github.com/dorianbrown/rank_bm25}} and set $k_1$=$1.5$ and $b$=$0.75$.
Results on dev sets are shown in Table~\ref{tab:bm25}. 
Different variants of \clocq are used for retrieving the KB subset,
where the focus is on larger initial $\mathcal{S}$ to measure the impact of BM25 (therefore the choice of a large $p$ of $10$k).
Answer presence for top-1000 facts is comparable to the initial answer presence; 
but
% However, we observed
a significant drop was observed when taking only the top-100 facts.
This indicates that the
% such
basic bag-of-words model in BM25 matching
falls short
% is not sufficient
for % more
complex questions.
However, an IR-based filter is a viable choice when the number of facts that can be
``consumed'' is budgeted.
% of facts is restricted to a specific number, it could be a valid choice.

\begin{table}
\caption{Effect of BM25 on verbalized KB facts.}
\vspace*{-0.3cm}
\newcolumntype{G}{>{\columncolor [gray] {0.90}}c}
    \resizebox*{\columnwidth}{!}{
    	\begin{tabular}{l G G c c}
    		\toprule
        		\textbf{Benchmark}	 & \multicolumn{2}{G}{\textbf{LC-QuAD2.0-CQ}}   & \multicolumn{2}{c}{\textbf{ConvQuestions-FQ}}  \\
        		\midrule
        % 		\textbf{\clocq}	 & \textbf{Answer}  & \textbf{Context}  & \textbf{Answer}  & \textbf{Context}  \\
        % 			    & \textbf{presence} & \textbf{size}     & \textbf{presence} & \textbf{size}\\
        	    \textbf{Method  $\downarrow$}    & \textbf{Ans. pres.}  & \textbf{$|\mathcal{S}|$}  & \textbf{Ans. pres.}  & \textbf{$|\mathcal{S}|$}  \\
                \toprule
                \textbf{\clocq} ($k$=1, $p$=10k)     &  $0.782$   &   $3.7$k &  $0.704$   &   $1.9$k \\
                \textbf{    + BM25} (top-100)        &  $0.625$   &   $0.1$k &  $0.509$   &   $0.1$k \\
                \textbf{    + BM25} (top-1000)       &  $0.726$   &   $0.8$k &  $0.630$   &   $0.7$k \\
                \midrule
                \textbf{\clocq} ($k$=3, $p$=10k)     &  $0.846$   &   $9.7$k & $0.775$   &   $5.9$k \\
                \textbf{    + BM25} (top-100)        &  $0.614$   &   $0.1$k &  $0.462$   &   $0.1$k \\
                \textbf{    + BM25} (top-1000)       &  $0.742$   &   $1.0$k &  $0.648$   &   $0.9$k \\
                \midrule
                \textbf{\clocq} ($k$=5, $p$=10k)         &  $0.872$   &   $15.1$k & $0.796$   &   $10.3$k \\
                \textbf{    + BM25} (top-100)        &  $0.605$   &   $0.1$k &  $0.456$   &   $0.1$k \\
                \textbf{    + BM25} (top-1000)       &  $0.747$   &   $1.0$k &  $0.648$   &   $0.9$k \\
    		\bottomrule
    	\end{tabular}
    }
    \label{tab:bm25}
    \vspace*{-0.3cm}
\end{table}

\subsection{Effect of fact-based KB indexing}

% In this section
Fact-centric KB storage is a foundation for \clocq: 
we now analyze its effect
% of our proposed KB index
on runtimes for search space reduction. 
% and compare with existing KB interfaces.
Our comparison points % compared with
are the available Wikidata SPARQL endpoint\footnote{\url{https://query.wikidata.org/bigdata/namespace/wdq/sparql?format=json}} ({\textsc{QueryService}}) and triple pattern queries issued to the {\textsc{Hdt}~\cite{fernandez2013binary}} KB interface.
We subtracted network latencies when measuring runtimes.
% of baselines.

\myparagraph{Basic functionalities}
Our first experiment was on the two basic functionalities required for KB-QA:
retrieving all facts of a given KB item (neighborhood), 
and measuring the
% KB
distance between two given KB items (KB-distance).
% and measuring the connectivity among two KB items (connectivity).
For baselines, we optimized the amount of required queries and implemented the distance checks as for \clocq (Sec.~\ref{sec:kb-storage}).
We took 1 million random KB items for the neighborhood lookups, and 1 million random KB item pairs for the connectivity checks.
Average runtimes (per KB item/KB item pair) are shown in Table~\ref{tab:kg-representation}.
We found that \textsc{Hdt} has a better performance than the Wikidata \textsc{QueryService}, making use of its efficient implementation via bit-streams.
However, \clocq can improve neighborhood lookups by a factor of $10$ and $10^3$
% , compared with
over
\textsc{Hdt}
and \textsc{QueryService}, respectively.
When measuring KB-distances, the effect becomes even larger: \clocq is $10^3$ and $10^4$ times faster than \textsc{Hdt}
and the \textsc{QueryService}, respectively.
The
% RAM
memory 
consumption for the \clocq KB index is slightly higher than that for \textsc{Hdt},
but this is still much lower than what loading the raw KB dump into memory would consume.
% but still feasible due to the memory-efficient representation of the KB index.

\myparagraph{Effect on search space reduction}
% To verify that our findings also hold on realistic data,
We now compare runtimes with these KB interfaces for search space reduction on the LC-QuAD dev set. 
While \clocq makes use of the neighborhood and KB-distance functions, only the neighborhood function is necessary in \textsc{Elq} and \textsc{TagMe}.
% of the KB
% is required.
We observe similar trends as before:
% in the previous experiment:
runtimes of \clocq are much better when using the \clocq KB index. The \textsc{QueryService} script
% could
did not
% even
terminate within a reasonable amount of time.
Interestingly, these trends also hold for \textsc{Elq} and \textsc{TagMe}: when using the \clocq KB index for search space reduction, the runtime is significantly reduced.
This shows that our fact-based KB index is valuable beyond this specific use in \clocq.
% The reduction is much larger for \clocq, as it takes KB-distances into account in its KB-aware scoring.

Gains in runtime are due to the fact-centric KB index, which is specifically designed
for providing efficient KB access for QA functionalities.
KB interface baselines may provide very fast
% efficient
KB access for general-purpose querying, 
but fall short for the more
specific
requirements in QA.
% KB functionalities required in QA. 

%but the issue generally calls for further research.
%
% - error-class 1: no answer-connecting item among candidate items\\
% - error-class 2: answer-conneting item not ranked properly (in top-1)\\
% class 1 (lcquad) : 44.8 \%\\
% class 2 (lcquad) : 55.2 \%\\
% class 1 (convquestions) : 46.7 \%\\
% class 2 (convquestions) : 53.3 \%\\

% % avoiding a table due to space shortage :(
% \myparagraph{Anecdotal examples} We analyzed several cases where \clocq was able to disambiguate to the right KB
% % item-tuple
% item-combination
% at the top-1 position, but none of the baselines could. We present two instances here: \utterance{What does the spouse of Thomas Müller, Lisa Trede-Müller, work as in 2018?} $\mapsto$ \struct{<spouse, Thomas Müller, Lisa, Lisa Müller, work, 2018>}; and, ii) \utterance{Who is the son of the brother of Queenie Padilla?} $\mapsto$ \struct{<nephew, brother, Queenie Padilla>}.
% % no space
% % ; and iii) \utterance{Write the long complex question 3 here?} $\mapsto$ \struct{<KB-ent-1, KB-pred-1, KB-ent-2, KB-pred-2>}.}
% % 2-3 correct question-KB tuple pairs where \clocq gets it in top-1 but not baselines.
% % What is the local dialing code for Melbourne?
% % Melbourne, 

% % what is p:
% % We apply this intuition as follows: when extracting the facts for some KG item, we test whether the number of facts in which the KG item does not appear in the subject position (object or qualifier object), is greater than some pruning threshold \textit{p}.

\begin{table} \small
    \caption{Comparison of KB interfaces w.r.t. functionalities.}
    \vspace*{-0.3cm}
    % \resizebox*{\columnwidth}{!}{
    	\begin{tabular}{l c c c}
    		\toprule
    		\textbf{KB interface}	 & \textbf{\textsc{QueryService}}  & \textbf{\textsc{Hdt}~\cite{fernandez2013binary}} & \textbf{\clocq}  \\ 
    		\toprule
    		\textbf{RAM consumed}       &	$-$ 	&	$220 GB$        &  $340 GB$	\\ \midrule
    		\textbf{Neighborhood}		&	$1.48 \times 10^{-2} s$ 	&   $6.73 \times 10^{-4} s$	&  $4.98 \times 10^{-5} s$   \\ 
    		\textbf{KB-distance}	    &	$2.46 \times 10^{-2} s$ 	&	$5.43 \times 10^{-3} s$	&  $3.23 \times 10^{-6} s$   \\	
    		\bottomrule
    	\end{tabular}%}
    % (basic functionalities).}
    \label{tab:kg-representation}
    \vspace*{-0.3cm}
\end{table}

\begin{table} \small
    \caption{Timing KB interfaces for search space reduction.}
    \vspace*{-0.3cm}
    % \resizebox*{\columnwidth}{!}{
    	\begin{tabular}{l c c c}
    		\toprule
    		\textbf{KB interface} & \textbf{\textsc{QueryService}}  & \textbf{\textsc{Hdt}~\cite{fernandez2013binary}}  & \textbf{\clocq} \\
    		% \textbf{Method} \textdownarrow         &  & \textbf{Runtime} \\ 
    		\toprule
    		\textbf{\clocq}                               &	$-$       &	$971$ s	&	$0.54$ s	\\
    		\textbf{\textsc{Elq}~\cite{li2020efficient}}		&	$0.89$ s	&	$0.62$ s  &	$0.12$ s  \\
    		\textbf{\textsc{TagMe}~\cite{ferragina2010tagme}}	    &	$19$ s	&	$1.25$ s	&	$0.52$ s	\\
    		\bottomrule
    	\end{tabular} %}
    \label{tab:kg-if-context}
    \vspace*{-0.3cm}
\end{table}

% \vspace*{-0.2cm}
\section{Discussion}
%Overall, we showed that
Disambiguating not only entities, but also general concepts, types, or predicates when establishing the search space,
% has 
%an impact on the upper bound for the downstream QA system.
% great benefits
is generally beneficial
for QA systems. This is something that is done by \clocq but is \textit{beyond NED} systems. 
%Further, we identified 
The detected trade-off between answer presence and search space size is an important factor:
increasing $|\mathcal{S}|$ improves answer presence but also injects noise, whereas a smaller search space could potentially be cleaner and easier to
% traverse
explore
by the QA system.
%one either brings in noise into the search space to enhance the answer presence, or relies on a smaller and cleaner set of facts with a lower upper bound.
This trade-off is closely connected to the choice of $k$
% , i.e. the number of KB items per question word, 
and the amount of facts for a specific KB item that is admitted into $\mathcal{S}$, that is controlled by our other parameter $p$.
%For QA, setting $k=1$ can inherently limit the downstream performance.

%%% PC: we want to put the dynamic-k setting as main proposal now,
%%% given that it does not need to be tuned and has an even stronger
%%% performance than a static value. Therefore, I adjusted the following text a bit
Among the static settings for $k$, we found $k\text{=}5$ to perform best on the considered benchmarks.  For other types of questions (e.g. simpler questions or list questions), 
the appropriate setting may have to be reconsidered.
% Overall, we recommend using a development/tuning set of withheld questions that are characteristic for an application or benchmark to determine good settings for the \clocq parameters.
The degree of ambiguity of question words 
% can also be
is a key factor:
we found a dynamic setting of $k$ (per question word) to perform the best among our variants.

%%%GW: following sentence is unclear (to me)
%While the upper bound of $~80\text{-}85\%$ can occur limiting at the first glance,
%we followed the standard approach for establishing the search space (for our baselines),
%and used GRAFT-Net for detecting the answer in our QA experiment.

%Since there is quite some gap between the QA performance and upper bound
%we argue that the benchmarks are quite challenging
%(there are no existing results, and we used a set of more complex questions from LC-QuAD2.0).
% Since there are no  (we extracted a set of more complex questions of LC-QuAD2.0),
%
%%%GW: now discuss the gap
% \GW{read this and adjust if needed!}
The answer presence obtained by \clocq lies in the range of 80 to 90 percent. This seems to indicate that downstream KB-QA methods cannot achieve a perfect
% recall
answering performance.
% for QA.
But on a practical note,
there is no QA system yet which gets anywhere near $100\%$ performance on realistic benchmarks. While state-of-the-art methods on some simpler benchmarks have reported a performance of $60 - 80\%$,
% show recall values around 90\% on their leaderboards,
the datasets of complex questions used in our experiments %, LC-QuAD 2.0 and ConvQuestions,
are much more demanding.
% as indicated by the poor performance of the state-of-the-art.
% \GW{maybe, another sentence with best known results from the literature ???}
In fact, we observe a substantial gap between the answer presence in the search space and the actual
%%% PC: performance is better, since we do not measure recall for QA (only P@1, MRR, H@5)
% recall
performance
of a state-of-the-art QA system.
% remove if space short
% Even scores for Hit@5 are well behind the upper bound.
% The main bottleneck is in the QA itself, 
% % not in the search space construction (anymore)
% %%% PC: if we say that the problem lies only in QA, this puts the paper in a weaker position,
% %%% since we do not target this problem really. Therefore, I added a connection of how we can help to improve.
% and a sophisticated method for search space contraction has potential to enhance performance of future QA systems.

\section{Related work}
\label{sec:rel}

\myparagraph{KB interfaces}
% In the past, most KB-QA methods accessed the KB via SPARQL queries.
Optimizing KBs for executing SPARQL queries is a well-studied problem~\cite{ erling2010virtuoso, gurajada2014triad, neumann2008rdf, urbani2016kognac, weiss2008hexastore}.  Urbani and Jacobs~\cite{urbani2020adaptive} recently proposed \textsc{Trident} for enabling different kinds of workloads (e.g. SPARQL, graph analytics) on large KBs. HDT~\cite{fernandez2013binary} encodes triples using bitmaps. It constructs two individual integer-streams holding predicates and objects, adjacent to some given subject, % are established. 
and two additional bit-streams for encoding connections between these predicates and objects. 
Due to multiple indexes, triple pattern queries can be answered very efficiently using HDT.
% Leeka et al.~\cite{leeka2016quark} propose to index quantifiable qualifiers,
% such as the confidence for a triple, 
% to enable efficient top-$k$ SPARQL querying.
% Nguyen et al.~\cite{nguyen2014don} assign each fact-specific predicate a unique ID,
% and represent the qualifiers in facts with this fact-specific predicate as subject, 
% to avoid the representation via intermediate nodes. 
% Despite these works on efficient KB encodings and interfaces~\cite{fernandez2013binary, leeka2016quark, urbani2016kognac, urbani2020adaptive}, there is still some gap between the provided functionalities and these required by KB-QA systems. 
These works focus on optimizing queries on triple stores. However, the problem of retrieving the complete facts of a KB item including qualifier information is a typical task in KB-QA, but is not targeted.

\myparagraph{Named entity disambiguation}
In named entity disambiguation (NED), the goal is to map entity mentions to the corresponding real-life concept:
pages in Wikipedia or entries in curated KBs like Wikidata.
\textsc{Tagme}~\cite{ferragina2010tagme} leverages Wikipedia anchors to detect entity mentions, looks up possible mappings, and scores these with regard to a collective agreement implemented by a voting scheme. In \textsc{Aida}~\cite{hoffart2011robust}, a mention-entity graph is established, and the mentions are disambiguated jointly by approximating the densest subgraph. More recently, van Hulst et al.~\cite{van2020rel} proposed a framework \textsc{Rel} for end-to-end entity linking, building on state-of-the-art neural components. \textsc{Elq}~\cite{li2020efficient} jointly performs mention detection and disambiguation leveraging a BERT-based bi-encoder.
These methods are optimized for computing the top-$1$ entity per mention, and mostly return only the top-ranked entity in the disambiguation. Top-$1$ NED is prone to errors that can propagate through the answering pipeline~\cite{shen2019multi,yih2015semantic}.
Early work in \textsc{S-Mart}~\cite{yang2015s}
% introduced structured multiple additive regression trees, and
applied statistical models of regression trees on a set of (mention, entity)-pairs and corresponding features. Unlike most other works, \textsc{S-Mart} returned top-$k$ disambiguations per mention. However, since it is proprietary, their code was not available for comparison.
% NED systems disambiguate, by definition, only mentions: they ignore references to predicates, general concepts, and types.
% While usage of NED in KB-QA has been the norm so far, we identify this as a major limitation for complex KB-QA.

\myparagraph{Search space reduction}
% In general,
Methods in complex KB-QA mostly follow one of two approaches~\cite{saharoy2021question}:
i) disambiguating entities, predicates, and types over the whole KB~\cite{ding2019leveraging,hu2018state,shen2019multi,vakulenko2019message}, for e.g., by leveraging question-word specific index lists~\cite{shen2019multi,vakulenko2019message} for subsequent semantic parsing; and, ii) applying NED as an initial step to focus the remaining computation on a restricted search space~\cite{bao2016constraint,bast2015more,bhutani2019learning,luo2018knowledge,qiu2020hierarchical,sawant2019neural,sun2019pullnet,xu2016question,yih2015semantic}.
The focus here is on improving the second line of work: instead of performing top-$1$ or top-$k$ NED, we disambiguate \textit{all} question cue words and compute the top-$k$ results per question token. This leads to a search space that is more likely to contain the relevant KB items \textit{and} the correct answer(s). % Entity and relation linking, e.g.
\textsc{Earl}~\cite{dubey2018earl} takes an approach of disambiguating both entity and predicate mentions. We generalize this direction by disambiguating all keywords in the question.

\section{Conclusions and future work}
\label{sec:confut}

We introduced answer search space reduction as a vital task in KB-QA. % , % and presented an efficient method for this task.
We showed that our proposal
\clocq,
based on the threshold algorithm over score-ordered lists
containing KB items with different relevance signals,
% and
is more successful
in retaining answers in its reduced search space
than 
% outperforms
a wide variety of general-purpose methods
for named entity disambiguation.
\newline{\empty}

\section*{Acknowledgments} We would like to thank Jesujoba Alabi (Max Planck Institute for Informatics, Germany) and Zhen Jia (Southwest Jiaotong University, China) for their help in preliminary experiments on integrating \clocq with KB-QA systems. We would also like to acknowledge Magdalena Kaiser (Max Planck Institute for Informatics, Germany) for her help in the KB cleaning process.
Part of this work was supported by the
ERC Synergy Grant ``imPACT'' (No. 610150).

% \clearpage

\balance

\bibliographystyle{ACM-Reference-Format}
\bibliography{clocq}

\end{document}